\documentclass[prx,superscriptaddress,aps, longbibliography, twocolumn, reprint, showpacs]{revtex4-2}
\usepackage[breaklinks=true,colorlinks=true,linkcolor=blue,urlcolor=blue,citecolor=blue]{hyperref}
\usepackage{bbm}

\usepackage[T1]{fontenc}
\usepackage[utf8]{inputenc}
\usepackage[dvipsnames]{xcolor}
\usepackage{amsmath}
\usepackage{amssymb}
\usepackage{graphicx}

\newcommand{\be}{\begin{equation}}
\newcommand{\ee}{\end{equation}}
\DeclareMathOperator{\diag}{diag}
\newcommand{\lambdaF}{\lambda_\text{F}}
\newcommand{\lambdaB}{\lambda_\text{B}}
\newcommand{\str}{\mathop{\rm str}}
\newcommand{\sdet}{\mathop{\rm sdet}}
\newcommand{\corr}[1]{\langle #1\rangle}
\newcommand{\Corr}[1]{\left\langle #1\right\rangle}
\newcommand{\Ei}{\mathop{\rm Ei}}
\renewcommand{\Re}{\mathop{\rm Re}}

\renewcommand{\hat}[1]{{\widehat #1}}

\begin{document}

\title{Sensitivity of (multi)fractal eigenstates to a perturbation of the Hamiltonian}

\author{M. A. Skvortsov}
\affiliation{L. D. Landau Institute for Theoretical Physics, Chernogolovka 142432, Russia}

\author{M.~Amini}
\affiliation{Department of Physics, University of Isfahan, Hezar Jerib, 81746-73441 Isfahan, Iran}

\author{V.~E.~Kravtsov}
\affiliation{Abdus Salam International Center for Theoretical Physics, Strada Costiera~11, 34151 Trieste, Italy}
\affiliation{L. D. Landau Institute for Theoretical Physics, Chernogolovka 142432, Russia}

\begin{abstract}
We study the response of an isolated quantum system governed by the   Hamiltonian drawn from the Gaussian Rosenzweig-Porter random matrix ensemble to a perturbation  controlled by a small parameter. We focus on the density of states, local density of states and the eigenfunction amplitude overlap correlation functions which are calculated exactly using the mapping to the supersymmetric nonlinear sigma model. We show that the susceptibility of eigenfunction fidelity to the parameter of perturbation can be expressed in terms of these correlation functions and is strongly peaked at the localization transition: It is independent of the effective disorder strength in the ergodic phase, grows exponentially with increasing disorder in the fractal phase and decreases exponentially in the localized phase. As a function of the matrix size, the fidelity susceptibility remains constant in the ergodic phase and increases in the fractal and in the localized phases at modestly strong disorder. We show that there is a critical disorder strength  inside the insulating phase such that for disorder stronger than the critical the fidelity susceptibility decreases with increasing the system size. The overall behavior is very similar to the one observed numerically in a recent work by Sels and Polkovnikov [Phys. Rev. E {\bf 104}, 054105 (2021)] for the normalized fidelity susceptibility in a disordered XXZ spin chain.
\end{abstract}

\maketitle

%%%%%%%%%%%%%%%%%%%%%%%%%%%%%%%%%%%%%%%%%%%%%%%%%%%%%%%%%%%%%%%%%%%%%%%%%%%%%%%%%%%%%%
\section{Introduction}
%%%%%%%%%%%%%%%%%%%%%%%%%%%%%%%%%%%%%%%%%%%%%%%%%%%%%%%%%%%%%%%%%%%%%%%%%%%%%%%%%%%%%%
Parametric statistics is a general term for the response of a disordered quantum system to the variation of
parameters of a Hamiltonian.
In the case of the classical Wigner-Dyson
random matrix theory (RMT) \cite{Mehta}
 one is usually interested in the {\it spectral response} to the parameter
variation,
the  best known
examples are the so called `level velocity' $dE_{n}/d\lambda$ and the `level curvature' $d^{2}E_{n}/d\lambda^{2}$ \cite{Thouless74,Wilk88,KrZirn92,Simons93,Szafer93,SimonsLeeAltshuler,Zakrewski93,FOppen94}, where $E_{n}(\lambda)$ is the $n$-th eigenvalue of the Hamiltonian
\be\label{H-1-2}
\hat{H}(\lambda)=\hat{H}_{1}\cos\lambda + \hat{H}_{2}\sin\lambda,
\ee
in which $\hat{H}_{1}$ and $\hat{H}_{2}$ are drawn from certain disorder ensembles and $\lambda$ is a parameter.
%%%%%%%%%%%%%%%%%%%%%%%%%%%%%%%%%%%%%%%%%%%%%%%%%%%%%%%%%%%%%%%%%%%%%%%%%%%%%%%%%%%%%%%%%%%%%%%%%%%%%%%%%%%%%%%%%%%
\begin{figure}[h!]
\centering
\includegraphics[width=0.8\linewidth]{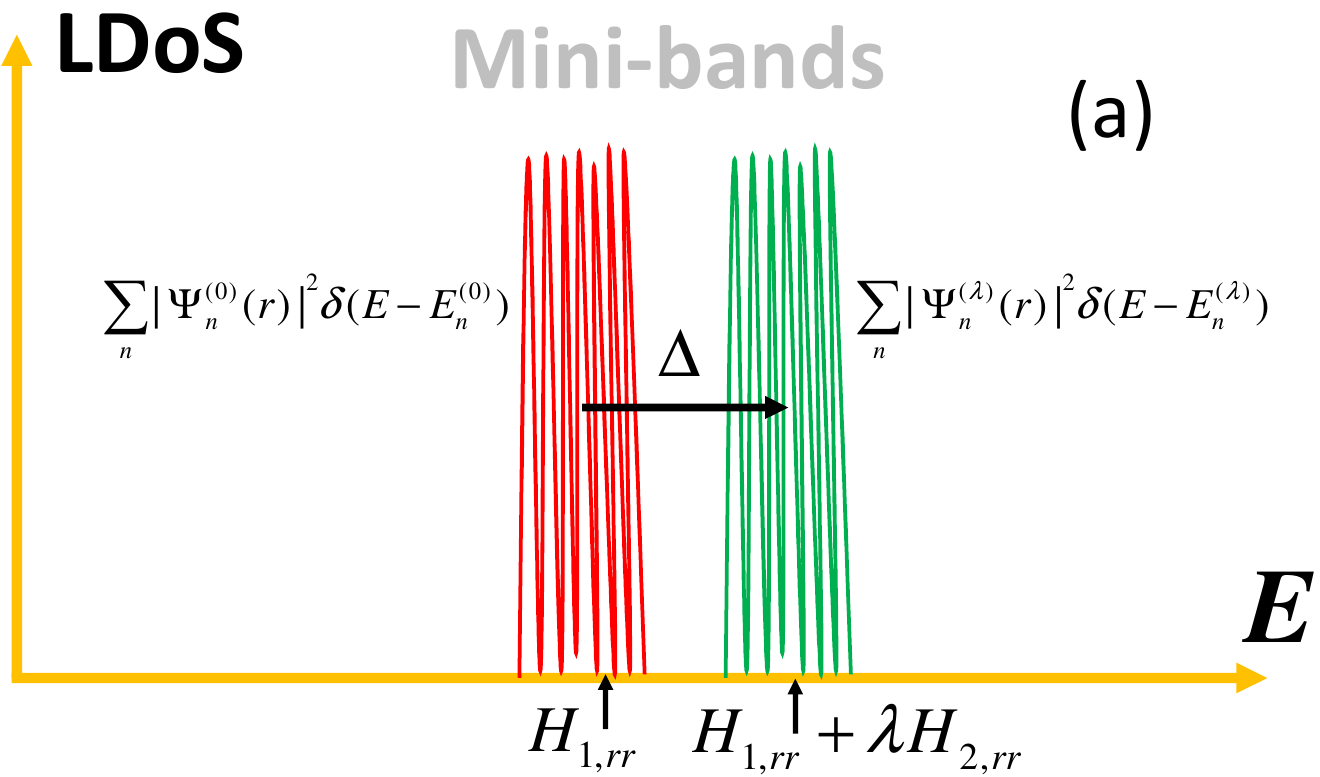}
\includegraphics[width=0.45\linewidth]{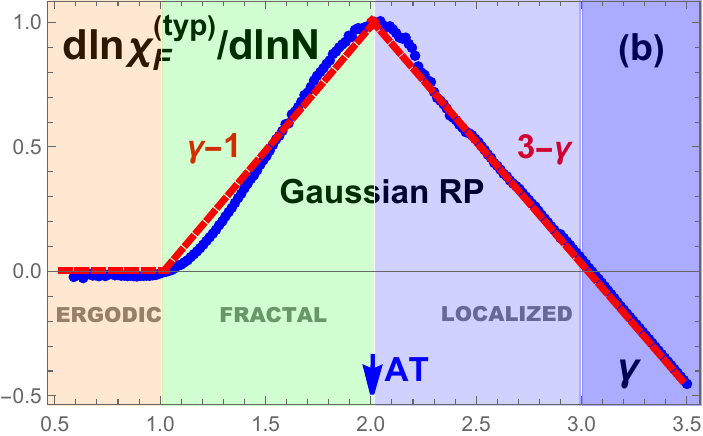}
\includegraphics[width=0.45\linewidth]{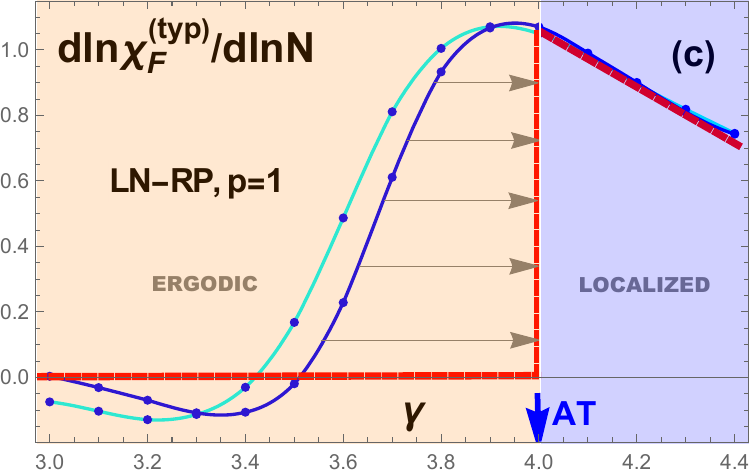} \\
\caption{{\bf Main results.}
{\bf (a)}~Shift of an entire mini-band of the width $\Gamma$ in the spectrum of LDoS consisting of a macroscopic number of states,   caused by a parametric shift of a  diagonal
matrix element $H_{rr}$ corresponding to the observation point $r$.
\textbf{(b)} The logarithmic derivative of the typical fidelity susceptibility $\chi_\text{F}^\text{(typ)}\sim K(\delta,0)$ as a function of the effective disorder parameter $\gamma$
[for the definition, see Eq.\ (\ref{gamma-def})]
for the Gaussian Rosenzweig-Porter (GRP) model in its ergodic, fractal and localized phases. Blue points are the results of exact diagonalization of
corresponding random matrices [shown is the discrete derivative $d\ln K(\delta,0)/d\ln N$ between $N=1024$ and $N=2048$].
The red, dashed, {\it symmetric}  with respect to the Anderson transition (AT) point $\gamma=2$ curve is the analytical prediction. The fractal phase at $1<\gamma<2$ manifests itself in the exponential growth $\propto N^{\gamma-1}$ of the typical fidelity susceptibility   as a function of $\gamma$ (and power-law growth as a function of $N$).  The fidelity susceptibility   has a maximum at  the localization transition at $\gamma=2$. In the localized phase  the susceptibility $\propto N^{3-\gamma}$ decreases as a function of $\gamma$ but as a function of $N$ it shows a peculiar behavior. It grows with $N$ for $2<\gamma<3$ and decreases with $N$ for $\gamma>3$. In the limit $N\rightarrow\infty$ it is infinite for $2<\gamma<3$ (light blue filling) and zero if $\gamma>3$ (dense blue filling).
Thus there exist two localized phases
with drastically different responses to a local perturbation.
The susceptibility in the ergodic phase is small and independent of $\gamma$.
\textbf{(c)} the logarithmic derivative $d\ln K(\delta,0)/d\ln N$
as a function of   $\gamma$ for the  Rosenzweig-Porter model with log-normal distribution of off-diagonal matrix elements (LN-RP) in the symmetric point $p=1$ which is equivalent to the Anderson localization model on random regular graph (RRG) \cite{LN-RP-RRG20,LN-RP-RRG21}. In this model the fractal phase does not exist. As the result, the typical fidelity susceptibility evolves with increasing $N$ to a {\it highly asymmetric} discontinuous curve shown by a red dashed line. The blue lines correspond to numerical diagonalization of random matrices drawn from LN-RP, $p=1$ ensemble with $N=4096$, 8192 (cyan) and $N=8192$, 16384 (blue) and taking the discrete derivative w.r.t.\ $\ln N$. Very similar curves are obtained for the Anderson model on RRG and for the Heisenberg spin-1/2 chain in a random field \cite{Tikh-Mir5}.}
\label{fig:main-result}
\end{figure}
%%%%%%%%%%%%%%%%%%%%%%%%%%%%%%%%%%%%%%%%%%%%%%%%%%%%%%%%%%%%%%%%%%%%%%%%%%%%%%%%%%%%%%%%%%%%%%%%%%%%%%%%%%%%%%%%%%%
\begin{figure}[tbh]
\center{
\includegraphics[width=0.8 \linewidth,angle=0]{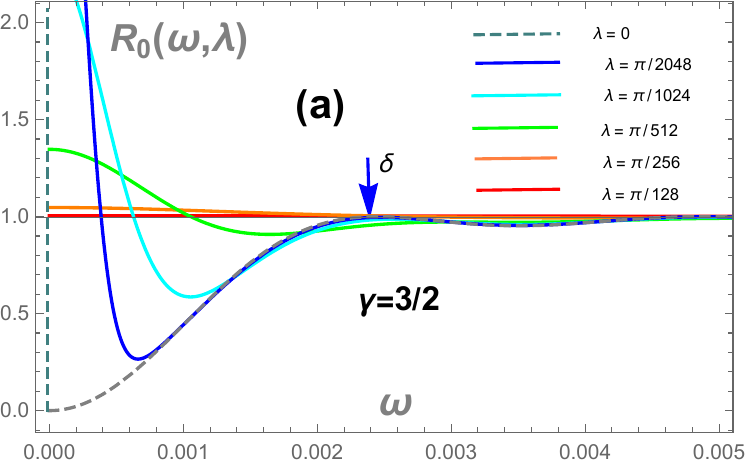}
\includegraphics[width=0.48 \linewidth,angle=0]{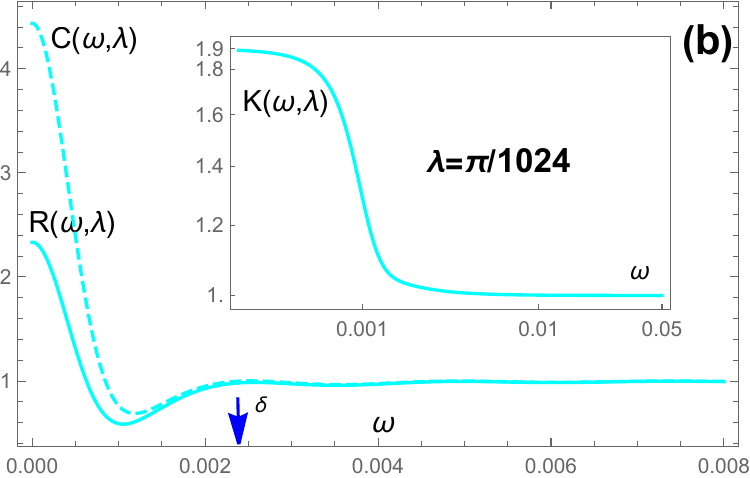}
\includegraphics[width=0.49 \linewidth,angle=0]{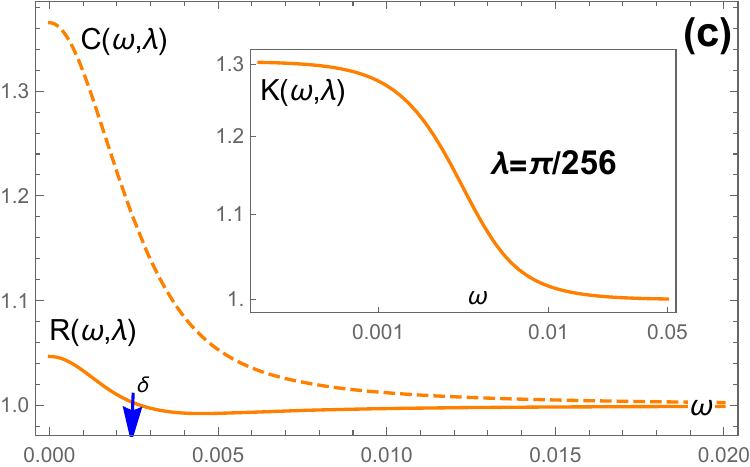}}
\caption{\textbf{Correlation functions} $R(\omega,\lambda)$, $C(\omega,\lambda)$, $K(\omega,\lambda)$ for the unitary GRP
ensemble in the fractal phase $\gamma=3/2$ at $\Gamma/\delta=32$ and $\delta=0.0024$ in the large-$N$ limit. The plots are obtained analytically
from Eqs.\ (\ref{R-lb}), (\ref{barC-int}), (\ref{int}).
{\bf (a)} Evolution of the DoS correlation function $R(\omega,\lambda)$ with increasing the parameter $\lambda$ in Eq.\ (\ref{H-1-2}).
\textbf{(b)} and \textbf{(c)}
The correlation functions  $R(\omega,\lambda)$, $C(\omega,\lambda)$, $K(\omega,\lambda)$ at $\lambda=\pi/1024$ and $\lambda=\pi/256$, respectively.
The $\delta$-peak of level self-correlation is progressively broadened   and the level repulsion correlation hole diminishes in $R(\omega,\lambda)$ and $C(\omega,\lambda)$ as $\lambda$ increases. The eigenfunction amplitude overlap (EAO) correlation function  $K(\omega,\lambda)$ demonstrates the broadened eigenfunction self-correlation peak rectified for the eigenvalue correlations. }
\label{fig:CRK}
\end{figure}
%%%%%%%%%%%%%%%%%%%%%%%%%%%%%%%%%%%%%%%%%%%%%%%%%%%%%%%%%%%%%%%%%%%%%%%%%%%%%%%%%%%%%%%%%%%%%%%%%%%%%%%%%%%%%%%%%%%

For the diffusive dynamics beyond the classical RMT the parametric density-of-states (DoS) correlation function and the level curvature distribution in the crossover between the orthogonal and the unitary ensemble was
studied in \cite{AndreevAltshuler95, KrYur96} using the formalism of the supersymmetric nonlinear sigma model (SUSY NLSM) \cite{Efetov_book}.
The dynamics of eigenvalues with $\lambda$, which plays a role of time, is  sensitive to the Anderson localization and is
 a generalization of the Dyson's Brownian motion \cite{Dyson62} of the gas of energy levels
(`Pechukas gas' \cite{Pechukas})
in the classical RMT. In the delocalized phase it exhibits the
avoided crossing of the `word lines' of energy levels as functions of $\lambda$ and in the localized phase the crossing is sharp.

The dynamics of the Pechukas gas is very peculiar  
if the wave functions are extended but not ergodic, obeying the multifractal statistics \cite{Mirlin-rev} (as e.g. in the point of Anderson transition). 
The connection between this dynamics  and the eigenfunction statistics \cite{ChalkerLerSmith} allowed to derive a non-trivial relationship between the level compressibility $\chi$ and the fractal
dimension $D_{q}$ of eigenfunctions \cite{ChalkerKrLer96}.

This incomplete review of old results shows that
quite a bit is known about the spectral  parametric statistics. However by now information on the parametric statistics of {\it eigenfunctions}, especially for the multifractal case, is still lacking.

At the same time,
the multifractal eigenfunction statistics appears to be quite ubiquitous. Not only
it is common to critical points of Anderson
transitions in non-interacting systems, a mounting evidence has emerged recently
\cite{Laflorencie19, Tarzia20} that the many-body
localized (MBL) phase in interacting systems  is a phase with
multifractal  statistics of eigenfunctions in the Hilbert space.  

Recently, the non-ergodic extended phase with fractal eigenfunction statistics
is shown \cite{RP15} to emerge   in the simplest extension of the Wigner-Dyson (WD) theory, the Gaussian Rosenzweig-Porter (GRP) random matrix ensemble
\cite{RP}.  Like in the classical RMT,  all matrix entries in this ensemble are Gaussian random  variables fluctuating independently of each other around zero with the variance that is identical for all the diagonal and all the off-diagonal matrix elements. The crucial difference is that the variance   of the off-diagonal matrix elements is parametrically smaller than that of the diagonal ones and it decreases as a certain power $N^{-\gamma}$ of the matrix size $N$, while the variance of diagonal entries is chosen to be 1. This breaks the basis-rotation invariance of the classical RMT and allows for the non-ergodic phases to emerge in the thermodynamic limit in the special  basis in which the model is formulated.

Despite its
quite specific formulation, this random matrix ensemble appears to have a predictive power even for
the MBL transition in interacting systems such as dirty bosons and spin chains \cite{Tarzia20}.
A particular case of this model arises also in the quantum random energy problem
\cite{SmelQREM20, QREM-FeigIoffe} which has
important applications in quantum computing \cite{SmelQREM20}.
The Rosenzweig-Porter model with logarithmically-normal distribution of off-diagonal
entries (LN-RP) \cite{LN-RP20} has a particularly rich two-parameter phase diagram \cite{LN-RP-RRG20,LN-RP-RRG21}. It is shown \cite{LN-RP-RRG21} that at a certain relationship between the typical value and the variance of the log-normal distribution LN-RP model belongs to the same universality class as the Anderson model on Random Regular Graph (RRG). The latter model is often considered as a toy model for MBL \cite{DeLuca2014, TMS2016, TikhMir_toy}.

Given a simplicity of the GRP model, in this paper we study the parametric statistics of local operators in this model with the goal to identify the generic features  intrinsic to all random systems in its multifractal phase.
%%%%%%%%%%%%%%%%%%%%%%%%%%%%%%%%%%%%%%%%%%%%%%%%%%%%%%%%%%%%%%%%%%%%%%%%%%%%%%%%%%%%%%%%%%%%%%%%%%%%%%%%%%%%%%%%%%%
\section{Target properties and review of main results}
%%%%%%%%%%%%%%%%%%%%%%%%%%%%%%%%%%%%%%%%%%%%%%%%%%%%%%%%%%%%%%%%%%%%%%%%%%%%%%%%%%%%%%%%%%%%%%%%%%%%%%%%%%%%%%%%%%%
Statistical properties of a quantum system are naturally described by the DoS and local DoS (LDoS) pair correlation functions:
 \begin{gather}
\label{R}
  R_E(\omega,\lambda)
  =
  \delta^{2}
  \biggl< \,
  \sum_{n,m}
  \delta(E-E_n^{(0)})
  \delta(E+\omega-E_m^{(\lambda)})
  \biggr> ,
\\
  C_E(\omega,\lambda)
  =
  (N\delta)^{2}
  \biggl< \,
  \sum_{n,m}
  \rho_n^{(0)}(r) \rho_m^{(\lambda)}(r)
  \hspace{24.5mm}
\nonumber
\\[-3pt]
  \hspace{26.3mm}
  {} \times
  \delta(E-E_n^{(0)})
  \delta(E+\omega-E_m^{(\lambda)})
  \biggr>
  ,
\label{C}
\end{gather}
where $N$ is the dimension of the Hilbert space (size of an $N\times N$ random matrix), $\delta$ is the mean level spacing at the band center,
$\rho_m^{(\lambda)}(r) = | \Psi_m^{(\lambda)}(r) |^2$ is the density operator, $\Psi_{n}^{(\lambda)}(r)=\langle r|n(\lambda)\rangle$ is the $n$-th eigenfunction of the Hamiltonian (\ref{H-1-2}) corresponding to the parameter $\lambda$ at a point $r$, and $E_{n}^{(\lambda)}$ is the corresponding eigenvalue. Angular brackets $\corr{\dots}$ stand for disorder average.

Since the dependence of the correlation functions \eqref{R} and \eqref{C} on the reference energy $E$ is very slow, one typically considers them at the band center. That will be implied throughout the text, so we will suppress the subscript $E$: $R(\omega,\lambda) \equiv R_0(\omega,\lambda)$ and $C(\omega,\lambda) \equiv C_0(\omega,\lambda)$.

While the DoS correlator $R(\omega,\lambda)$ is determined only by the eigenvalues, the LDoS correlator $C(\omega,\lambda)$ also encodes information about the eigenfunctions.
In many cases, the spectrum and eigenfunctions are uncorrelated. The best known examples are the classical RMT \cite{Mehta} (fully-ergodic states) and random diagonal matrices (perfectly localized states).
Then the function $C(\omega,\lambda)$ reproduces level repulsion features of the DoS correlator $R(\omega,\lambda)$ and it is instructive to focus on their ratio \cite{CueKrav07}:%
\be
\label{K_C_R}
K(\omega,\lambda) = \frac{C(\omega,\lambda)}{R(\omega,\lambda)}.
\ee
We will show that for the GRP model $K(\omega,\lambda)$, indeed, coincides, in its principal detail, with the correlation function of the density operators
\be
\label{tilde-K}
 \tilde{K}(\omega,\lambda) = N^{2}\langle \rho_{m}^{(0)}(r)\,\rho_{n}^{(\lambda)}(r)\rangle,
\quad \,\,
E_n-E_m=\omega
\ee
(both $E_n$ and $E_m$ taken close to the band center).
For this reason, the correlation function $K(\omega,\lambda)$ will be referred to as the eigenfunction amplitude overlap (EAO) correlation function.

In this paper we consider three types of perturbations $\hat{H}_{2}$ in Eq.\ (\ref{H-1-2}):
(i) the diagonal drive when $\hat{H}_{2}$ is the diagonal part of the RP random matrix,
(ii) the off-diagonal drive when $\hat{H}_{2}$ is the off-diagonal part of this matrix, and
(iii) the local density drive $\hat{H}_{2}=(N\delta)\,|r\rangle\,\langle r|$.

The main effect of the \emph{diagonal drive} is a random shift $\Delta$ of the entire mini-band \cite{return} of the nearly-resonant states in the local spectrum following the shift of the matrix element $H_{r,r}$ [see Fig.\ \ref{fig:main-result}(a)].
In the case of the \emph{off-diagonal drive} the mini-band does not move as a whole.  However, in both cases (i) and (ii) the individual levels do move. At a small perturbation strength $\lambda$ the shift of a single level manifests itself in $K(\omega, \lambda)$ as the broadening of the self-correlation $\delta$-peak in $K(\omega,0)$. The second effect of perturbation is the suppression of level repulsion seen as a correlation hole in $C(\omega,\lambda)$ and $R(\omega,\lambda)$ at $\omega\lesssim \delta$, see Fig.\ \ref{fig:CRK}.

The point of a special focus in this work is the {\it fidelity susceptibility} \cite{GFS07,GFS15,Sels21}
characterizing the overlap of the bare and perturbed wave functions averaged over all states:
\be\label{chiF}
\chi_\text{F}
=
- \partial_{\lambda^{2}}
\frac1N \sum_n
\bigl|\langle\Psi_{n}^{(0)}|\Psi_{n}^{(\lambda)}\rangle\bigr|^{2}
_{\lambda=0} .
\ee
For a generic perturbation $\hat{H}_{2}$, the fidelity susceptibility bears information on quantum phases.
We consider the specific case of the local density drive
 $\hat{H}_{2}=(N\delta)\,|r\rangle\,\langle r|$
 when the respective $\chi_\text{F}$ is insensitive to phases of eigenfunctions and can be expressed via the correlation function $C(\omega,0)$. We show that the \emph{typical} fidelity susceptibility for the local density drive in the Gaussian RP (GRP) ensemble is given by
\be
\label{fidel-Fin}
\chi_\text{F}^\text{(typ)} \sim C(\delta,0) \sim K(\delta,0) .
\ee

One of the main results of the paper is the exponential growth of the typical fidelity susceptibility $\chi^\text{(typ)}_{F}\sim N^{\gamma-1}$ as a function of the ``disorder parameter'' $\gamma$ in the fractal non-ergodic extended (NEE) phase ($1<\gamma<2)$. This result is derived analytically and confirmed numerically. In the ergodic phase ($\gamma<1$) the susceptibility $\chi_\text{F}^\text{(typ)}$ is independent of $\gamma$ and is much smaller than in the NEE phase, while in the localized phase ($\gamma>2$) it exponentially decreases as $N^{3-\gamma}$ with increasing $\gamma$, see Fig.\ \ref{fig:main-result}(b). The maximum of $\chi_\text{F}^\text{(typ)}(\gamma)$ in the limit of large $N$ is located exactly at the localization transition $\gamma=2$.

An important feature of the dependence of $\ln\chi_\text{F}^\text{(typ)}$ on $\gamma$  in the GRP model is a symmetric character of this dependence with respect to the Anderson transition point $\gamma=2$, see Fig.\ \ref{fig:main-result}(b). It is instructive to compare this behavior with the one for the log-normal RP model \cite{LN-RP20} which is equivalent to the Anderson localization model on RRG \cite{LN-RP-RRG20,LN-RP-RRG21}. In that case the NEE phase is only a finite-size effect, which manifests itself in the \emph{asymmetric} shape of the dependence of $\ln\chi_\text{F}^\text{(typ)}$ on $\gamma$, see Fig.\ \ref{fig:main-result}(c). With increasing the matrix size $N$ the asymmetry grows and in the limit of an infinite system one obtains a discontinuous curve for $d\ln\chi_\text{F}^\text{(typ)}/d\ln N$ vs.\ $\gamma$ shown by the red dashed line in Fig.\ \ref{fig:main-result}(c). A very similar behavior has been obtained in Ref.\ \cite{Tikh-Mir5}
for the EAO correlation function of neighboring eigenstates in RRG and in the random-field spin-1/2 Heisenberg chain (cf.\ right panels of Figs.\ 3 and 6 in Ref.\ \cite{Tikh-Mir5}).

One should specially mention the scaling of the typical fidelity susceptibility $\chi_\text{F}^\text{(typ)}$  with the system size $N$. The susceptibility is power-law divergent with $N$ in the multifractal ($1<\gamma<2$) and in the mildly localized  ($2<\gamma<3$) phases,   albeit with different exponents $\gamma-1$ and $3-\gamma$. However, it  vanishes in the limit $N\rightarrow\infty$ in the strongly localized phase for $\gamma>3$, see Fig.\ \ref{fig:main-result}(b) and Fig.\ \ref{fig:chi-sketch}.

A recent detailed study of an XXZ spin-1/2 chain in a random field \cite{Sels21} revealed a picture (see Fig.\ 1 therein) which is qualitatively similar to our Figs.\ \ref{fig:main-result}(b) and \ref{fig:chi-sketch}.
For small disorder the   typical   fidelity susceptibility is almost size-independent and slowly-varying with increasing disorder, for intermediate disorder it gets enhanced (blowing up with increasing the system size), reaches a peak and then falls down at larger disorder. In the latter region there is an apparent fixed point $W^{*}$ in the disorder strength $W$ (similar to $\gamma^{*}=3$ in our model)
such that with increasing the system size the fidelity susceptibility grows at $W<W^{*}$ and decreases at $W>W^{*}$.

We present a qualitative picture explaining the two drastically different regimes of $N$-dependence of the typical fidelity susceptibility both in the Rosenzweig-Porter model with long-range hopping and in short-range hopping models (including interacting systems with two-body interaction corresponding to short-range hopping in the Hilbert space). In the Rosenzweig-Porter model such a behavior is due to the non-exponential localization of typical eigenstates, while in short-range systems it is due to the competition between typical exponentially localized states and rare Mott's pairs of states \cite{Mott68,Mott70,Ivanov2012} emerging as the result of hybridization of two resonant exponentially localized states.

\begin{table}
\caption{{\bf The principal characteristics of the Gaussian RP model in the three phases:} the bandwidth $E_\text{BW}$,
the mean level spacing $\delta$, and the width  $\Gamma$ of the Lorentzian  in Eq.\ (\ref{Kom-Lorenz}) (the latter in the fractal phase has the meaning of the mini-band width).
Both $\delta$ and $\Gamma$ are calculated at the band center.
}
\label{T:GRP}
\begin{ruledtabular}
\begin{tabular}{cccc}
phase & ergodic & fractal & localized \\
\hline
& $\gamma < 1$ & $1<\gamma<2$ & $2<\gamma$ \\
condition & $W/\sigma \ll \sqrt{N}$ & $\sqrt{N} \ll W/\sigma \ll N$ & $N \ll W/\sigma$ \\
$E_\text{BW}$ & $2\sqrt{N} \sigma$ & $\sim W$ & $\sim W$ \\
$\delta$ & $\sim \sigma/\sqrt{N}$ &   $\sqrt{2\pi} W/N$ &   $\sqrt{2\pi} W/N$
\\
$\Gamma$ & $\sim\sqrt{N} \sigma$ &  $\sqrt{2\pi} N\sigma^2/W$ & $  \sim \sigma$ \\
\end{tabular}
\end{ruledtabular}
\end{table}

%%%%%%%%%%%%%%%%%%%%%%%%%%%%%%%%%%%%%%%%%%%%%%%%%%%%%%%%%%%%%%%%%%%%%%%%%%%%%%%%%%%%%%
\section{ The model and its main properties \label{sec:Def }}
%%%%%%%%%%%%%%%%%%%%%%%%%%%%%%%%%%%%%%%%%%%%%%%%%%%%%%%%%%%%%%%%%%%%%%%%%%%%%%%%%%%%%%

The Rosenzweig-Porter model \cite{RP} is a generalization of the Wigner-Dyson random matrix model to the case when diagonal and off-diagonal elements fluctuate with different strengths.
Several types of the Rosenzweig-Porter ensembles with various distribution functions of off-diagonal matrix elements have been considered:
Gaussian \cite{RP15}, logarithmically-normal \cite{LN-RP20},
and power-law (Levy) \cite{BirTar_Levy-RP}.

In the simplest Gaussian RP (GRP) model, the Hamiltonian is an $N\times N$ Hermitian real symmetric (orthogonal ensemble, $\beta=1$) or complex (unitary ensemble, $\beta=2$) random matrix. Its entries are independent Gaussian-distributed random variables with zero mean and variances ($m\neq n$)
\be\label{variances}
  \corr{H_{nn}^2} = W^2,
\qquad
  \corr{|H_{nm}|^2} = \sigma^2.
\ee
Parameters $W$ and $\sigma$, together with the matrix size $N$, completely specify the GRP model.
In the limit of vanishing $W$, it is equivalent to the usual RMT, with all the states perfectly mixed (ergodic regime) and the average DoS given by the Wigner semicircle with the bandwidth $E_\text{BW}^\text{RMT} = 2\sqrt{N} \sigma$. In the limit of vanishing $\sigma$, we have a perfect localization at each site and the DoS follows the distribution of $H_{nn}$ with the dispersion $W$.

Remarkably, in the GRP model transition from the ergodic to the localized regime goes through an intermediate non-ergodic extended (NEE) phase with a  fractal statistics of eigenvectors \cite{RP15}. With increasing $W$, the crossover from the RMT to the NEE regimes takes place at $W\sim E_\text{BW}^\text{RMT}$.
At larger $W$, the off-diagonal disorder is too weak to mix all the states and only a part of them, $M\sim(\Gamma/W)N\sim\Gamma/\delta$, participates in the formation of new states from the localized orbitals. Here $\Gamma$ is the width of the mini-band given by (energy close to the band center implied)
\be
\label{Gamma-NEE}
  \Gamma^\text{NEE}
  =
  \sqrt{2\pi} \,\frac{N\sigma^2}{W}= 2\pi
	\,\frac{\sigma^{2}}{\delta}.
  \ee
As $W$ increases, the number $M$ of occupied sites in a typical eigenvector decreases and becomes of the order of 1 when $\Gamma\sim \delta$, where $\delta\sim W/N$ is the mean level spacing. This point marks the localization transition.

The relevant energy scales discussed above in the ergodic, fractal and localized regimes are summarized in Table \ref{T:GRP}.
Note that in the localized phase the mini-band collapses to a single level, so formally the mini-band width is zero. Yet the parameter $\Gamma$ in the numerator of Eq.\ (\ref{Kom-Lorenz}) remains finite, $\Gamma\sim \sigma\ll \delta$, but apparently looses the meaning of the mini-band width \cite{return}.

In order to study the GRP model in the thermodynamic limit, $N\to\infty$, it was suggested in Ref.\ \cite{RP15} to use the following scaling of off-diagonal matrix elements:
\be\label{gamma-def}
  \sigma^2 \propto N^{-\gamma} ,
\ee
while keeping $W$ independent of the matrix size $N$. Notice that so defined $\gamma$ plays a role of the {\it effective disorder parameter} determining the ratio $ W^{2}/(N\sigma^{2})\propto N^{\gamma-1}$ of the disorder potential to the kinetic bandwidth.

In the thermodynamic limit the crossovers between the phases discussed above at a finite $N$
turn into sharp phase transitions at $\gamma=1$ and $\gamma=2$, with the ergodic, fractal and localized phases corresponding to $\gamma<1$, $1<\gamma<2$ and $2<\gamma$, respectively (see Table \ref{T:GRP}).
The mini-band width obeys the following scaling \cite{return}:
\be\label{Gamma-all}
  \Gamma
  \propto
  \begin{cases}
    N^{(1-\gamma)/2} , & \text{ergodic;} \\
    N^{1-\gamma} ,     & \text{fractal.} \\
    % N^{-\gamma/2} ,    & \text{localized.} \\
  \end{cases}
\ee
In terms of the eigenvector statistics, the ergodic phase ($\gamma<1$) corresponds to uniform spread of the eigenfunction over all sites, while in the localized phase ($\gamma>2$) the eigenfunction is highly peaked on a single site. In the intermediate NEE phase ($1<\gamma<2$), the eigenvector is fractal with an extensive number of occupied sites which, however, constitutes a vanishing fraction of all available sites. This fractal behavior can be characterized with the help of the eigenvector support set defined as a set of sites $r$, which is sufficient for the normalization condition $\sum_{r\in\text{SS}}|\Psi_{n}(r)|^{2}=1-\epsilon$ to be fulfilled with any prescribed accuracy $\epsilon\ll 1$. The volume of the support set in the fractal phase scales with the matrix size as $M\sim N^{D}$, with the fractal dimension \cite{RP15}
\be
  D = 2-\gamma
\ee
restricted by $0<D<1$. The sites from the support set are nearly resonant and form a compact  {\it mini-band} of levels in the energy space of the width $\Gamma\propto N^{D-1}$ [see Eq.\ \eqref{Gamma-all}] that vanishes in the limit $N\rightarrow\infty$ (see Fig.~\ref{fig:main-result}).
%%%%%%%%%%%%%%%%%%%%%%%%%%%%%%%%%%%%%%%%%%%%%%%%%%%%%%%%%%%%%%%%%%%%%%%%%%%%%%%%%%%%%%%%%%%%%%%%%%%%%%%%%%%%%%%%%%
\begin{figure}[tb]
\centering
\includegraphics[width=0.8\linewidth]{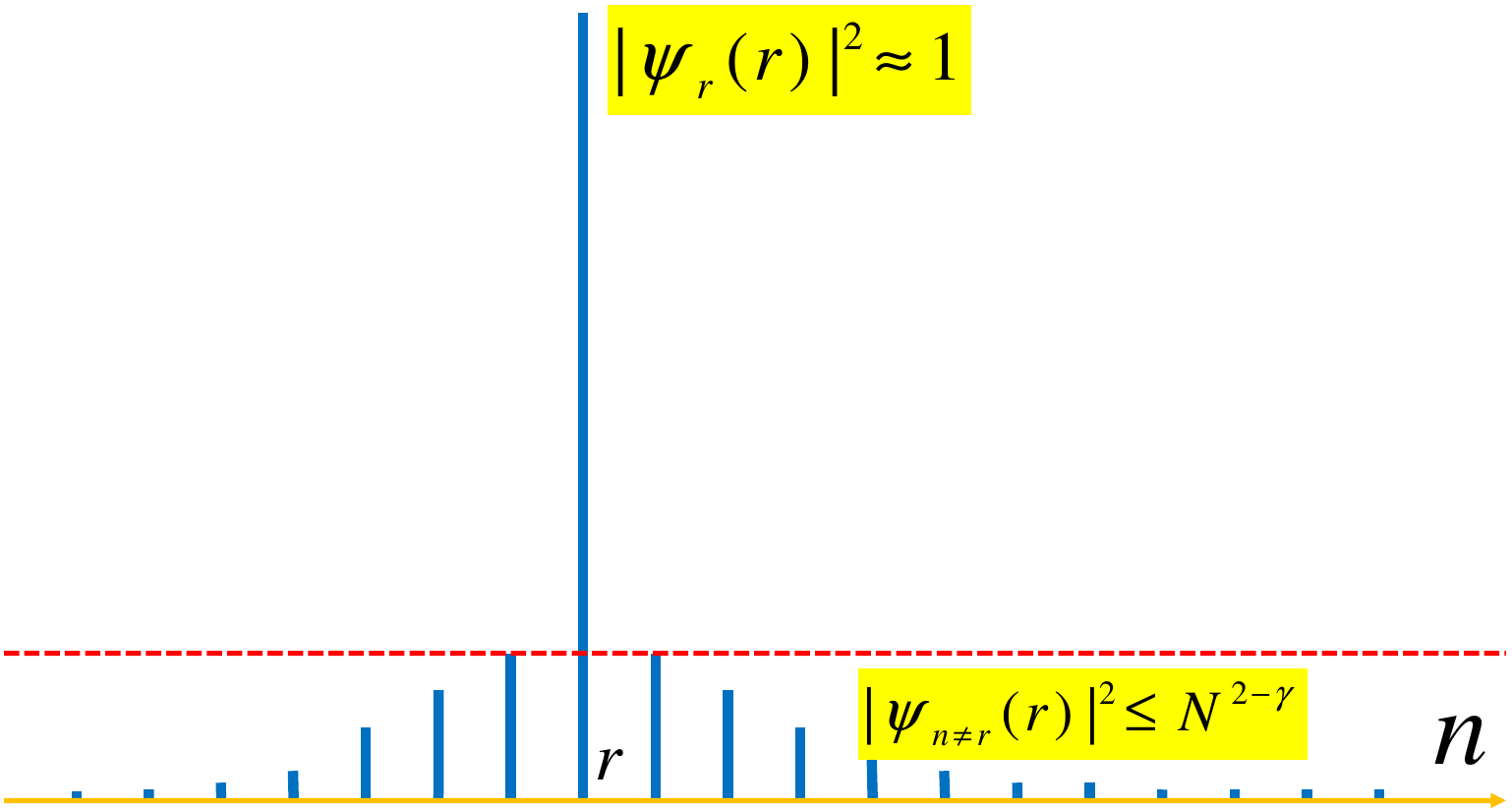}
\caption{\textbf{A typical localized state in the GRP model.} In the thermodynamic limit ($N\rightarrow\infty$) the localization radius is strictly zero.}
\label{fig:zero_r_loc}
\end{figure}
%%%%%%%%%%%%%%%%%%%%%%%%%%%%%%%%%%%%%%%%%%%%%%%%%%%%%%%%%%%%%%%%%%%%%%%%%%%%%%%%%%%%%%%%%%%%%%%%%%%%%%%%%%%%%%%%%%%
The eigenfunction amplitude statistics in the ergodic phase ($\gamma<1$) follows the Porter-Thomas distribution $P_\text{PT}(x) \propto x^{(\beta-2)/2}\exp(-cx^2)$, like in the classical RMT \cite{Mehta}. The same remains true in the NEE phase as well, provided one restricts $\Psi_{n}(r)$ to the support set \cite{BogomolnyRP2018}.

In general, the amplitude of an eigenvector $|\Psi_{n}(r)|^{2}=|\langle r|n \rangle|^2$ in the GRP model is well approximated by the \emph{Monthus surmise} introduced in Ref.\ \cite{Monthus17b}:
\be
\label{Bogomol}
  |\Psi_n(r)|^2 = \frac{|H_{nr}|^2}{(E_n-d_r)^2+(\Gamma/2)^{2}} .
\ee
Here $H_{nr}$ is the off-diagonal matrix element, $d_r=H_{rr}$ is the diagonal matrix element, and $E_n$ is an exact eigenvalue. In the ergodic phase and in NEE phase at the support set, the first term in the denominator of the Lorentzian \eqref{Bogomol} can be neglected, and one readily recovers the Porter-Thomas distribution.

The Monthus surmise (\ref{Bogomol}) can also be made meaningful in the localized phase ($\gamma>2$), if one identifies $\Gamma/2=\langle |H_{nm}|^{2}\rangle^{1/2} \sim N^{-\gamma/2}$ \cite{return}. In this case a wave function $\Psi_{n}(r)$ is strongly peaked at $n=r$ (see Fig.\ \ref{fig:zero_r_loc}), $E_n\approx d_n$ and Eq.\ \eqref{Bogomol} gives
$|\Psi_n(r)|^2 \sim \sigma^2/(d_n-d_r)^2$ unless $d_n$ and $d_r$ are in resonance with the accuracy of $\sigma$. Since the difference $d_n-d_r$ is typically larger than $\delta$, the amplitude of the wave function $\Psi_{n\neq r}(r)$ is typically bounded from above by
\be\label{one-site}
  |\Psi_{n\neq r}(r)|^2 \lesssim \sigma^2/\delta^2 \propto N^{2-\gamma} .
\ee
In fact, for the majority of sites with $|d_n-d_r|\sim E_\text{BW} \propto N^{0}$ the wave function is much smaller: $|\Psi_{n}(r)|^2\sim\sigma^2/W^2\propto N^{-\gamma}$. The contribution of $n\neq r$ sites to the normalization integral can be estimated as $\sum_{n\neq r} |\Psi_n(r)|^2 \sim \sum_{n\neq r} \sigma^2/\delta^2(n-r)^2 \sim \sigma^2/\delta^2 \propto N^{2-\gamma}$ due to fast convergence of the sum, which effectively saturates by a finite number of levels $n$ adjacent to $r$. Hence,%
\be\label{psi-r-r}
1-|\Psi_r(r)|^2
\sim
\sigma^2/\delta^2 \propto N^{2-\gamma},
\ee
 and in the thermodynamic limit ($N\rightarrow\infty$) $\langle r|n\rangle
=\delta_{nr}$, i.e. an eigenvector is localized strictly on a single site at \emph{any}\ $\gamma>2$, as shown in Fig.\ \ref{fig:zero_r_loc}.

An important extension of the GRP is the logarithmically-normal RP ensemble (LN-RP) with the {\it tailed} distribution of off-diagonal matrix elements $P(|H_{nm}|)$:
\be\label{LN-dist}
P(x)
=
\frac{A}{x}
\exp \left[-\frac{\ln^2(x/x_\text{typ})}{2p \ln(1/x_\text{typ})}\right].
\ee
The LN-RP ensemble is specified by two parameters: $\gamma$ and $p$. The former determines the {\it typical} value of the off-diagonal matrix element:
\be
H^\text{(typ)}\equiv\exp\langle \ln |H_{nm}| \rangle \sim N^{-\gamma/2}.
\ee
The parameter $p$ controls the strength of the {\it tail} in the distribution function Eq.\ (\ref{LN-dist}), which becomes thicker as $p$
increases. In the limit $p\rightarrow 0$ the LN-RP model approaches the GRP, while the special
choice $p=1$ brings the model in the same universality class as the Anderson model on Random Regular Graph
\cite{LN-RP-RRG21}. Remarkably, for $p>0$ the LN-RP model possesses a non-trivial dynamics. For instance the ensemble-averaged {\it survival probability} in some parameter region decays with time   as a {\it stretch exponent} \cite{LN-RP-RRG21}, in contrast to a simple exponential decay in the delocalized phases of GRP.

One can also consider \cite{BirTar_Levy-RP} the Levy-RP model with a power-law distribution $P(x)=A/x^{1+k}$ (the parameter $k>0$) truncated at $|x|<x_\text{typ}$, which is in many respects similar to LN-RP model.

%%%%%%%%%%%%%%%%%%%%%%%%%%%%%%%%%%%%%%%%%%%%%%%%%%%%%%%%%%%%%%%%%%%%%%%%%%%%%%%%%%%%%%%%%%%%%%%%%%%%%%%%%%%%%%%%%%%%%%%%%%%%%%%%
\section{Fidelity susceptibility $\chi_\text{F}$ and eigenstate amplitude overlap correlation function $K(\omega)$ \label{sec: F-K }}
%%%%%%%%%%%%%%%%%%%%%%%%%%%%%%%%%%%%%%%%%%%%%%%%%%%%%%%%%%%%%%%%%%%%%%%%%%%%%%%%%%%%%%%%%%%%%%%%%%%%%%%%%%%%%%%%%%%%%%%%%%%%%%%%

\subsection{General expression for $\chi_\text{F}$}

The fidelity of two states $|n(0)\rangle$ and $|n(\lambda)\rangle$ belonging to the Hamiltonians $\hat{H}(0)$ and $\hat{H}(\lambda)$ is defined as
\be\label{fidel}
F^{2}_{n}(\lambda) = |\langle n(0)|n(\lambda)\rangle|^{2} .
\ee
Since the set of eigenstates $|m\rangle=|m(0)\rangle$ is complete and orthonormal, one can write
\be
  F_{n}^{2}(\lambda)
  =
  1 - \sum_{m\neq n} |\langle m|n(\lambda)\rangle|^2.
\ee

In order to express the fidelity behavior at $\lambda\to0$ in terms of the eigensystem of the Hamiltonian $\hat{H}(0)$, we employ the Hellmann-Feynman relation:
\be
\langle m|\partial_{\lambda} n\rangle =  \frac{\langle m|\partial_{\lambda}\hat{H}|n\rangle}{E_{n}-E_{m}} ,
\ee
which can be derived by differentiating the Schr\"odinger equation $\hat{H}(\lambda)|n(\lambda)\rangle = E_n(\lambda)|n(\lambda)\rangle$ with respect to the parameter $\lambda$.
The fidelity susceptibility defined by Eq.~\eqref{chiF} is then given by \cite{GFS07,GFS15,Sels21}
\be\label{fidel-gen}
  \chi_\text{F}^{(n)}
  \equiv
  - \frac{dF_{n}^{2}}{d\lambda^{2}}\bigg|_{\lambda=0}
  =
  \sum_{m\neq n}
  \frac{|\langle m|\partial_{\lambda}\hat{H}|n \rangle|^{2}}{(E_{m}-E_{n})^{2}}
  \bigg|_{\lambda=0}.
\ee
Equation (\ref{fidel-gen}) is general and applies to any system.

Due to the denominator $(E_m-E_n)^2$ in Eq.\ \eqref{fidel-gen}, the fidelity susceptibility $\chi_\text{F}^{(n)}$ is a random and strongly fluctuating function of the state $n$. To get rid of this irregularity,
 one can also define a fidelity susceptibility   averaged over the spectrum
{\it in any given realization} of disorder:
\be\label{spectral-av}
\chi_\text{F}=\frac{1}{N} \sum_{n}\chi_\text{F}^{(n)}=\frac{1}{N} \sum_{n}\sum_{m\neq n}
  \frac{|\langle m|\partial_{\lambda}\hat{H}|n \rangle|^{2}}{(E_{m}-E_{n})^{2}}
  \bigg|_{\lambda=0}.
	 \ee
In what follows we will study this {\it spectral-averaged} fidelity susceptibility.

%%%%%%%%%%%%%%%%%%%%%%%%%%%%%%%%%%%%%%%%%%%%%%%%%%%%%%%%%%%%%%%%%%%%%%%%%
\subsection{Fidelity susceptibility for a local density drive}
%%%%%%%%%%%%%%%%%%%%%%%%%%%%%%%%%%%%%%%%%%%%%%%%%%%%%%%%%%%%%%%%%%%%%%%%%%

For further applications we choose a specific perturbation $\partial_{\lambda}\hat{H}$ proportional to the {\it local density} operator:%
\be\label{projector}
\partial_{\lambda}\hat{H}=N\delta\,|r\rangle \langle r|.
\ee
Then the matrix element in Eq.\ \eqref{fidel-gen} becomes a product of two wave functions at the given point $r$:
\be
\label{mdHn}
\langle m|\partial_{\lambda}\hat{H}|n \rangle=N\delta\,
\langle m|r\rangle\,\langle r|n\rangle \equiv N\delta\,\Psi_{m}^{(0)}(r)\Psi^{(0)*}_{n}(r) .
\ee
As a result,   the spectral-averaged   fidelity susceptibility acquires the form
\be\label{chi-int}
\chi_\text{F} = 2\delta \int_{0}^{\infty} \frac{{\mathfrak C}(\omega,r)}{\omega^{2}}\,d\omega,
\ee
where
\be
{\mathfrak C}(\omega,r)
=N\delta
\sum_{n}\sum_{m\neq n} \rho_n^{(0)}(r) \rho_m^{(0)}(r)
  \delta(\omega-\omega_{mn})
\ee
and $\omega_{nm}=E_n^{(0)}-E_m^{(0)}$.
Note that the {\it ensemble-averaged} value of ${\mathfrak C}(\omega,r)$ can be expressed in terms of $C_{E}(\omega,\lambda)$ defined in Eq.\ (\ref{C}):
\be\label{C-int}
\langle {\mathfrak C}(\omega,r)\rangle =\int \frac{dE}{N\delta}\, C_{E}(\omega,\lambda=0)\sim C(\omega,0).
\ee
Since $C_{E}(\omega,\lambda=0)$ is weakly $E$-dependent inside the spectral band-width $|E|\lesssim N\delta $ and is almost zero beyond it, the integral in Eq.\ (\ref{C-int}) is equal to $C(\omega,0)\equiv C_0(\omega,0)$ up to a prefactor of order 1.

The fidelity susceptibility given by Eq.\ \eqref{chi-int} is a random quantity with some distribution function $P(\chi_\text{F})$. An important feature of this distribution is the presence of a power-law tail at large $\chi_\text{F}$, originating from the factor $\omega^2$ in the denominator of Eq.\ \eqref{chi-int}. Indeed, in this regime the sum    in Eq.\ (\ref{spectral-av}) is dominated by a single term with the smallest $\omega_{nm}=|E_{n}-E_{m}|\ll \delta$, and $P(\chi_\text{F})$ is determined by the strength of  level repulsion. With ${\mathfrak C}(\omega,r) \sim \omega^\beta$, we obtain the large-$\chi_\text{F}$ asymptotics: $P(\chi_\text{F})\sim\chi_\text{F}^{-(3+\beta)/2}$. This power-law tail causes the divergence of the {\it ensemble-averaged} fidelity susceptibility in the localized phase ($\beta=0$) and in the extended phase for the orthogonal symmetry class ($\beta=1$) (the extended phase of the unitary symmetry with $\beta=2$ is characterized by a finite $\corr{\chi_\text{F}}$).

The maximum of the distribution function $P(\chi_\text{F})$ is reached when {\it many} terms start to be relevant in
the sum in Eq.\ (\ref{fidel-gen}). Thus the {\it typical} value $\chi_\text{F}^\text{(typ)}$ of $\chi_\text{F}$,  characterized by the maximal  probability, corresponds to the smallest $|E_{n}-E_{m}|$ being of the order of the typical level spacing (that for RP models coincides with the mean level spacing $\delta$). This suggests that the typical fidelity susceptibility is given by:
\be\label{chi-typ-C}
\chi_\text{F}^\text{(typ)}
\sim
\delta\int_{\delta}^{\infty} C^\text{(typ)}(\omega,0) \,\frac{d\omega}{\omega^{2}},
\ee
where $C^\text{(typ)}(\omega,0)$ is the typical value of 
${\mathfrak C}(\omega,r)$:
\be\label{typ-C}
C^\text{(typ)}(\omega,\lambda=0) 
= 
\exp \langle \ln {\mathfrak C} (\omega,r)\rangle .
\ee

Taking into account that for $\omega>\delta$ the DoS correlation function $R^\text{(typ)}(\omega,0)=R(\omega,0)\approx 1$ [where $R^{(\text{typ})}(\omega,0)$ corresponds to the typical DoS correlation function similar to Eq.\ (\ref{typ-C})], one may rewrite Eq.\ (\ref{chi-typ-C}) as
\be\label{chi-typ-K}
\chi_\text{F}^\text{(typ)}
\sim
\delta\int_{\delta}^{\infty}  K^\text{(typ)}(\omega,0)\,\frac{d\omega}{\omega^{2}}.
\ee
Now, given that the integral in Eq.\ (\ref{chi-typ-K}) is dominated by
 the lower limit,
we arrive at
\be\label{chi-typ-fin}
\chi_\text{F}^{(\text{typ})}\sim K^{(\text{typ})}(\delta,0).
\ee
Equation (\ref{chi-typ-fin}) applies to a generic case when $K^{(\text{typ})}(\omega,\lambda)$ does not coincide with $K(\omega,\lambda)$, e.g.\ to the LN-RP model and for models with short-range hopping. For the Gaussian RP model with the tail-free distribution of off-diagonal matrix elements, $K^{(\text{typ})}(\omega,\lambda)$ coincides with $K(\omega,\lambda)$, and Eq.\ ({\ref{chi-typ-fin}) leads to the estimate~\eqref{fidel-Fin}.

\subsection{Parametric density correlation function at a diagonal drive within the Monthus surmise approximation}
\label{SS:K-approx}

In this subsection we make a simple derivation of the  parametric density correlation function $\tilde{K}(\omega,\lambda)$ [Eq.\ \eqref{tilde-K}] based on the Monthus surmise approximation, Eq.\ (\ref{Bogomol}). An exact derivation in Sec.\ \ref{sec: Dos_LDoS-analyt} based on the supersymmetric nonlinear sigma model confirms the validity of this simple derivation in the fractal phase at the energy scale $|\omega+\lambda W|\gg\delta$,
when level repulsion effects can be neglected.

In both derivations we assume the diagonal drive,
\be\label{H2}
  \hat{H}_{2} = \hat{d'} = \diag\{ d'_{n}\} .
\ee

Let us first consider the \emph{fractal} phase with the hierarchy of scales $\delta\ll\Gamma\ll E_\text{BW}$, see Table \ref{T:GRP}. Substituting Eq.\ \eqref{Bogomol} into Eq.\ \eqref{tilde-K}, we can express $\tilde{K}(\omega, \lambda)$ in terms of the eigenfunction overlap at $E_{n}-E_{m}=\omega$:
\be\label{L-L}
  \tilde{K}(\omega, \lambda)
  =
  \frac{4\pi^2N^2}{\Gamma^2} \left\langle |H_{nr}|^{2}|H_{mr}|^{2} \,
{\cal L}( d_{r}-\tilde{\omega} )\,{\cal L}( d_{r})\right\rangle,
\ee
where ${\cal L}(\omega)=(\Gamma/2\pi)[\omega^{2}+(\Gamma/2)^{2}]^{-1}$ is a normalized Lorentzian.

Equation \eqref{L-L} should be averaged over the matrix elements $H_{nr}$, $H_{mr}$, and $H_{rr}=d_r$ of the Hamiltonian $\widehat H_1$, and over the diagonal matrix elements $d'_r$ of $\widehat H_2$ (which determine the shifted frequency $ \tilde\omega = \omega - \lambda d'_r$), all of them are independently fluctuating.
The averaging over off-diagonal matrix elements gives $\sigma^4$, while the averaging over $d_r$ is Gaussian with the variance $W^2$ [Eq.\ \eqref{variances}].
Then using the fact that the convolution of two Lorentzians is a Lorentzian of the double width and relying on the relations of Table \ref{T:GRP}, one obtains:
\be
\label{Kom-Lorenz}
\tilde K(\omega,\lambda)
=
\frac{1}{\pi}
\frac{N}{M}
\left\langle\frac{\Gamma^{2}}{({ \omega - \lambda d'_n})^2+\Gamma^2}\right\rangle_{d'_{n}},
\ee
where $\langle\dots\rangle_{d'_n}$ denotes the averaging over the diagonal matrix element $d'_n$ of the Hamiltonian $\widehat H_2$, and $M=\Gamma/\delta$ is the number of states in the mini-band.
Note that the diagonal drive $\hat{d'}$ in Eq.\ \eqref{Kom-Lorenz} should not necessarily be Gaussian.

Equation \eqref{Kom-Lorenz} remains valid in the \emph{localized} phase characterized by  $M=1$ and $\Gamma\sim \sigma\ll \delta$ (see Table \ref{T:GRP}). Since the Monthus surmise is based on the Wigner-Weisskopf approximation \cite{Wigner-Weisskopf} and the latter is valid only for times shorter than the Heisenberg time $t_\text{H}=\hbar/\delta$, Eq.\ (\ref{Kom-Lorenz}) is justified only for $|\tilde{\omega}-\lambda d'_{n}|\gg \delta$. Therefore, in the localized phase with $\sigma\ll\delta$ the relaxation rate $\Gamma$ in the \emph{denominator} of Eq.\ (\ref{Kom-Lorenz}) is not well defined and should be neglected. Then Eq.\ (\ref{Kom-Lorenz}) acquires a perfectly perturbative form
\be\label{Kom-pert}
\tilde{K}(\omega,\lambda)
\propto
\frac{\sigma^{2}}{(\omega - \lambda d'_n)^2}
\qquad
\text{for $|\omega - \lambda d'_n|\gg \delta$},
\ee
as it should be with $\sigma$ being the smallest parameter of the problem.

By the order of magnitude, Eq.\ (\ref{Kom-Lorenz}) remains valid also in the \emph{ergodic} phase when $M\sim N$, $\Gamma\sim E_\text{BW}\to\infty$ and, consequently, $\tilde{K}(\omega,\lambda)\sim 1$.

%%%%%%%%%%%%%%%%%%%%%%%%%%%%%%%%%%%%%%%%%%%%%%%%%%%%%%%%%%%%%%%%%%%%%%%%%%%%%%%%%%%%%%%%%%%%%%%%%%%%%%%%%%%%%%%%%%%
\begin{figure}
\center{
\includegraphics[width=0.8 \linewidth,angle=0]{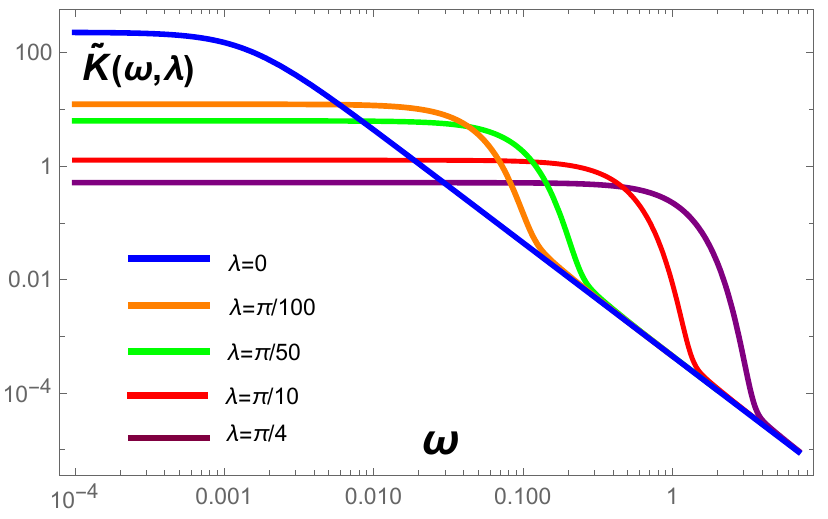}
}
\caption{\textbf{Density correlation function $\tilde{K}(\omega,\lambda)$ [Eq.\ (\ref{Kom-Lorenz})]}
for $\Gamma=0.0014$ and Gaussian distribution of $d'_n$ with the variance $W^2=1/2\pi$ at various values of the control parameter~$\lambda$.}
\label{fig:K-Gamma}
\end{figure}
%%%%%%%%%%%%%%%%%%%%%%%%%%%%%%%%%%%%%%%%%%%%%%%%%%%%%%%%%%%%%%%%%%%%%%%%%%%%%%%%%%%%%%%%%%%%%%%%%%%%%%%%%%%%%%%%%
For the Gaussian distribution of $d'_{n}$ the density correlation function given by Eq.\ (\ref{Kom-Lorenz}) is plotted in Fig.\ \ref{fig:K-Gamma}.

Vice versa, if the density correlation function $\tilde{K}(\omega,\lambda)$
is known, then using Eq.\ (\ref{Kom-Lorenz}) one can extract the effective distribution function ${\cal P}_{\lambda}(\Delta)$ of the random shift $\Delta=\lambda d'_{n}$ from the following relation:
\be\label{P-Delta}
{\cal P}_{\lambda}(\Delta)=\int \frac{dt}{2\pi}\,\frac{G(t,\lambda)}{G(t,0)}\,e^{i t \Delta},
\ee
where $G(t,\lambda)$ is the Fourier-transform of $K(\omega,\lambda)$.

Equation \eqref{P-Delta} is strictly derived under an assumption of the Lorentzian-like form of the density correlator \eqref{Kom-Lorenz}. Nevertheless, we will use it as an operational definition of ${\cal P}_{\lambda}(\Delta)$ in a generic situation.

\subsection{$\chi_\text{F}^\text{(typ)}$ and its scaling in different phases}

Here, based on Eqs.\ (\ref{fidel-Fin}), (\ref{Kom-Lorenz}) and (\ref{Kom-pert}), we discuss the fidelity susceptibility and its scaling in the limit $N\to\infty$.
Summarizing the results of Sec.\ \ref{SS:K-approx} for $\tilde K(\delta,0)$, plugging them into Eq.\ \eqref{fidel-Fin}, and using relations of Table \ref{T:GRP}, we obtain:
\be
\label{scaling-K-delta}
\chi_\text{F}^\text{(typ)}
\sim
\begin{cases}
   1 \propto N^{0}, & \gamma<1;  \\
   N\delta/\Gamma \propto N^{\gamma-1}, & 1<\gamma<2; \\
   N(\sigma/\delta)^2 \propto N^{3-\gamma}, & 2<\gamma.
\end{cases}
\ee

In order to illustrate the validity of Eq.\ (\ref{fidel-Fin}) in the localized phase, one may substitute Eq.\ (\ref{mdHn}) into Eq.\ (\ref{spectral-av}) and notice that the main contribution comes from $m=r$ (the choice $n=r$ doubles the result) when $\Psi_r(r)\approx1$. Then using the Monthus surmise (\ref{Bogomol}) for $\Psi_n(r)$ one obtains:
\be
\chi_\text{F} = 2 N \delta^2 \sum_{n\neq r}\frac{|H_{nr}|^{2}}{(E_{r}-E_{n})^{4}}.
\ee
For the tail-free Gaussian distribution of $H_{nr}$ the typical value of the fidelity susceptibility, $\chi_\text{F}^\text{(typ)}$, corresponds to the typical minimal distance $|E_{r}-E_{n}|\sim \delta\sim W/N$, while $|H_{nr}|^{2}$ can be replaced by its average $\sigma^2$. Thus we readily reproduce the last line in Eq.\ (\ref{scaling-K-delta}) confirming the validity of  Eq.\ (\ref{fidel-Fin}) in this particular case.

Taking the derivative of $\ln\chi_\text{F}^\text{(typ)}$ with respect to $\ln N$ eliminates $N$-independent prefactors   in Eq.\ (\ref{scaling-K-delta}), leading to the following asymptotically exact scaling:
\be
\label{log-der-chi}
  \frac{d\ln\chi_\text{F}^\text{(typ)}}{d\ln N}
  =
\begin{cases}
  0, & \gamma<1;  \\
  \gamma-1, & 1<\gamma<2; \\
  3-\gamma, & 2<\gamma.
\end{cases}
\ee
This result is shown by the red dashed line in Fig.\ \ref{fig:main-result}(b).

Note that the derivative $d\ln\chi_\text{F}^\text{(typ)}/d\ln N$ changes sign at a fixed point $\gamma^{*}=3$ which lies {\it inside} the localized phase. This result can be traced back to the behavior of the eigenfunction amplitude, Eq.\ (\ref{one-site}). Indeed, since for the GRP model $K(\omega,\lambda)\approx \tilde{K}(\omega,\lambda)$, both being independent of $r$, it follows from Eq.\ (\ref{tilde-K}) that
\be\label{anomal}
K(\delta,0)\approx  N\sum_{r}\left\langle|\Psi_{m}(r)|^{2} |\Psi_{m+1}(r)|^{2}\right\rangle .
\ee
The leading contribution to Eq.\ (\ref{anomal}) comes either from $r=m$ [where $\Psi_{r}(r)\approx 1$] or from $r=m+1$. Then Eqs.\ (\ref{fidel-Fin}) and (\ref{one-site}) provide an estimate
\be\label{cond-fixedpoint}
  \chi_\text{F}^\text{(typ)} \sim N\sigma^2/\delta^2 \propto N^{3-\gamma}.
\ee

Thus the change of sign of the
logarithmic
derivative $d\ln\chi_\text{F}^\text{(typ)}/d\ln N$ happens when the amplitude $|\Psi|^{2}$ of one state at a site where the adjacent in energy state is localized is of the order of $1/N$. In other words, at the fixed point $\gamma=\gamma^*$ the maximal overlap
\be\label{overlap}
J_{nm} = W\sum_{r}|\Psi_{n}(r)|^{2}|\Psi_{m}(r)|^{2}
\ee
between two different \emph{typical} localized states becomes of the order of $\delta$:
\be\label{fixed-point_cond}
 \max_m
 \,\{J_{n\neq m}\}\big|_{\gamma=\gamma^*} \sim \delta ,
\ee
with $\max_m\{J_{n\neq m}\}\gg\delta$ for $\gamma<\gamma^*$ (weaker disorder) and $\max_m\{J_{n\neq m}\}\ll\delta$ for $\gamma>\gamma^*$ (stronger disorder).

%%%%%%%%%%%%%%%%%%%%%%%%%%%%%%%%%%%%%%%%%%%%%%%%%%%%%%%%%%%%%%%%%%%%%%%%%%%%%%%%%%%%%%%%%%%%%%%%%%%%%%%%%%%%%%%%%%
\subsection{Generalization to short-range hopping models: the role of resonant Mott's pairs}
%%%%%%%%%%%%%%%%%%%%%%%%%%%%%%%%%%%%%%%%%%%%%%%%%%%%%%%%%%%%%%%%%%%%%%%%%%%%%%%%%%%%%%%%%%%%%%%%%%%%%%%%%%%%%%%%%%%

The condition \eqref{fixed-point_cond} for the critical disorder strength where $d\ln\chi_\text{F}^\text{(typ)}/d\ln N$ changes its sign was derived for the GRP model characterized by an infinite-range hopping.
In fact, it is much more general and applies also to systems with \emph{short-range} hopping, such as the Anderson model on $d$-dimensional lattices and graphs. However, in that case the meaning of $\max\{\cdot\}$ needs to be specified.

Indeed, a typical localized state in such systems has a single localization center (``head''), with $|\Psi_n(r)|^{2}$ decreasing exponentially as $r$ is shifting away from this center. Therefore two such states with a distance $R_{nm}$ between
their ``heads''
have an exponentially small overlap $J_{nm}\sim W \exp(-R_{nm}/\xi)$,
where $\xi$ is the localization radius. Since the typical distance between the localization centers $R_{nm}$ is of the order of the system size $L$, such ``single-headed'' states make an {\it exponentially small} contribution to $K(\delta,0)$ and hence to the fidelity susceptibility that would \emph{decrease} with increasing $L$.
This statement is true provided that only typical states are considered.

On the other hand, it is known \cite{Mott68,Mott70,Ivanov2012} that two {\it resonant} localized ``single-headed'' states $\Psi_n(r)$ and $\Psi_m(r)$
may effectively hybridize to form a Mott's pair of bonding and anti-bonding states, $\Psi_{\pm}(r)=[\Psi_n(r)\pm \Psi_m(r)]/\sqrt{2}$.
The resulting ``double-headed'' states have the level splitting $\Delta E\sim J_{nm}$ given by the overlap of the parent states $\Psi_n(r)$ and $\Psi_m(r)$. The distance $R_{nm}=R_{\Delta E}$ between the ``heads'' of the parent states required to form a Mott's pair grows with decreasing $\Delta E$ as $R_{\Delta E}\sim \xi\ln(W/\Delta E)$.
Though such pairs of ``double-headed'' states are rare, their overlap
\be
  J_{nm} \approx \frac{W}{4}
  \sum_{r} \left\{ |\Psi_{n}(r)|^{4}+|\Psi_{m}(r)|^{4} \right\},
\ee
is independent of the system size.
Thus the formation of Mott's pairs results in an \emph{exponentially large} gain in the overlap compared to that of the parent states.

This gain makes the Mott's pairs a serious competitor to ``single-headed'' states for the leading contribution to $K(\delta,0)$, with the winner depending on the particular type of the average considered. Mott's pairs contribution determines the \emph{far tail} of the distribution of random overlaps $J_{nm}$ which dominates in the \emph{mean} $K(\delta,0)$.
However, typical states are ``single-headed''. Therefore taking the \emph{typical average} $K^\text{(typ)}(\delta,0)$ [as in Eq.\ (\ref{chi-typ-K})] one cuts the tail of the distribution and strongly discriminates the contribution of Mott's pairs. As the result, at strong enough disorder [analogous to our $\gamma>\gamma^*$] the \emph{typical} $K^\text{(typ)}(\delta,0)\sim \chi^\text{(typ)}_{F}$ \emph{decreases} exponentially with the system size, in contrast to the \emph{mean} $K(\delta,0)$ which always (except for the case of one-dimensional single-particle localization) \emph{increases} with $L$. The reason for such an anomalous behavior is the dominance of Mott's pair contribution in the mean $K(\delta,0)$.

Indeed, it is known that  if only the Mott's pair contribution is taken into account, the averaged $K(\omega,0)$ would increase with decreasing $\omega\gtrsim\delta$ \cite{CueKrav07,Tikh-Mir5} and hence $K(\delta,0)$ would increase with increasing the system size $L$. Notice that only the pairs with the level splitting $\Delta E=\delta$ make a contribution to this quantity. Notice also that $R_{\Delta E=\delta}\propto\ln(W/\delta)\sim \ln N$ is growing with increasing the system size. Since the probability of forming a resonant Mott's pair of states  is proportional to the area $S=S(\Delta E)$ of a sphere of radius $R=R_{\Delta E}$ in a $d$-dimensional ($S\sim R^{d-1}$) or in an ultra-metric ($  S\sim \,{\cal K}^{R}$) space, this probability for $\Delta E=\delta\sim 1/N$ is increasing with the system size. As the overlap $J_{nm}$ for a Mott's pair is   independent of the system size, the increase of probability to find a Mott's pair leads to an increase of the averaged $K(\delta,0)$.

Summarizing, we can say that if the Mott's pair contribution is dominant in $K^\text{(typ)}(\delta,0)\sim \chi^\text{(typ)}_{F}$, the typical fidelity susceptibility grows  with the system size, otherwise it decreases with it.

The condition for the boundary between the two regimes can be formulated in a form of Eq.\ (\ref{fixed-point_cond}), with the left-hand side determined by the competition between ``double-headed'' and ``single-headed'' states. Since the probability to find a resonant Mott's pair decreases with increasing the level splitting $\Delta E$, there should be a maximal $\Delta E$ for which the contribution of the Mott's pairs to the typical $K^\text{(typ)}(\delta,0)$ still dominates over that of typical ``single-headed'' states. A particular value of $\Delta E$ is system-dependent, and it is the corresponding $\max_m\{J_{n\neq m}\}\sim \max\{\Delta E\}$ that enters the criterion (\ref{fixed-point_cond}) in the case of the short-range hopping systems.

The peculiarity of the GRP model is that in this case the localization is {\it not exponential}, contrary to models with short-range hopping. Therefore, even \emph{typical} ``single-headed'' states may lead to both increasing and decreasing with the systems size \emph{mean} $K(\delta,0)$ [which in the GRP model is of the same order as $K^\text{(typ)}(\delta,0)\sim \chi_\text{F}^\text{(typ)}$]. Another important difference is that in GRP the transition between the two regimes of diverging and vanishing fidelity susceptibility is a sharp phase transition in the limit $N\to\infty$ rather than a crossover, as in systems with short-range hopping.

%%%%%%%%%%%%%%%%%%%%%%%%%%%%%%%%%%%%%%%%%%%%%%%%%%%%%%%%%%%%%%%%%%%%%%%%%%%%%%%%%%%%%%
\section{Analytical results for parametric correlation functions
in~the case of diagonal drive}
\label{sec: Dos_LDoS-analyt}
%%%%%%%%%%%%%%%%%%%%%%%%%%%%%%%%%%%%%%%%%%%%%%%%%%%%%%%%%%%%%%%%%%%%%%%%%%%%%%%%%%%%%%

\subsection{General expressions in the fractal phase}

In this section we review analytical results for the DoS, LDoS and EAO parametric correlation functions obtained in the framework of the supersymmetric nonlinear sigma model (NLSM) for the \emph{unitary} GRP model in its \emph{fractal phase} subject to the \emph{diagonal drive}. The validity of the NLSM is justified by the large ratio $g=E_\text{Th}/\delta$ of the Thouless energy $E_\text{Th}$ to the mean level spacing that allows to neglect the longitudinal models and enforce the sigma-model constraint $Q^{2}=1$. In the fractal phase, the Thouless energy is given by $E_\text{Th}=\sqrt{\Gamma\delta}$ \cite{return} and hence $g=\sqrt{\Gamma/\delta}\gg1$, justifying the NLSM approach in the whole fractal phase where $\Gamma\gg\delta$ (see Table \ref{T:GRP}). Adopting the scaling Eq.\eqref{gamma-def} one obtains $g\propto N^{1-\gamma/2}\to\infty$ as long as $1<\gamma<2$.

The results we present below correspond to the limit $N\rightarrow\infty$ and apply to the fractal phase ($\Gamma\gg\delta$). They are calculated at the band center ($E\ll E_\text{BW}$ and $\omega\ll E_\text{BW}$) and are valid for an arbitrary value of $\omega/\delta$.
The general expressions for $R(\omega,\lambda)$ and
\be
\label{CC}
  C(\omega,\lambda)
  =
  \frac{N\delta}{\pi\Gamma}
  \bar{C}(\omega,\gamma),
\ee
as they emerge from the formalism of the NLSM, read:
\begin{gather}
\label{gen-R}
R(\omega,\lambda)
=
1
+\frac{1}{2}
  \Re \!
  \int_{-1}^{1} d\lambdaF \!
  \int_{1}^{\infty} \! d\lambdaB
\, P_R \, e^{S}
+
\frac{C(\omega,\lambda)}{N} ,
\\
\label{gen-C}
  \bar{C}(\omega,\lambda)
  =
  \Corr{ \frac{\Gamma^{2}}{\tilde{\omega}^{2}+\Gamma^{2}} }
  _{\!\!d'}
  +
  \Re \!
  \int_{-1}^{1} d\lambdaF \!
  \int_{1}^{\infty} \! d\lambdaB
  \frac{P_C \, e^{S}}{\lambdaB-\lambdaF}\,
  ,
\end{gather}
where integration is performed over two Cartan variables $\lambdaF$ and $\lambdaB$ specific to the unitary (class A) symmetry, the action is given by
\be
\label{action-final}
  S =
  \frac{i\pi}{\delta}(\lambdaB-\lambdaF)
  \Corr{ \frac{\Gamma \tilde{\omega}}{\tilde{M}_{B}^{1/2}} }
  _{\!\!d'}
  ,
\ee
and the functions $P_R(\lambdaB)$ and $P_C(\lambdaB)$ are defined as
\begin{gather}
\label{PR}
P_R
=
  \Corr{ 
  \frac{\Gamma^{2}(\Gamma-i\tilde{\omega}\lambdaB)}{\tilde{M}_{B}^{3/2}} 
  }
  _{\!\!d'}
^{\!2} ,
\\
\label{PC}
  P_C
  =
  \Corr{ \frac{2\tilde{M}_{B}\lambdaB+3i\Gamma\tilde{\omega}(\lambdaB^{2}-1)}{2 \tilde{M}_{B}^{5/2}/\Gamma^{3}} 
  } 
  _{\!\!d'}
  .
\end{gather}
In these expressions
\be
  \tilde{M}_{B}
  =
  \Gamma^{2}-\tilde{\omega}^{2}-2i\Gamma\tilde{\omega}\lambdaB ,
\ee
with $\tilde{\omega}=\omega-\lambda d'$ being a $\lambda$-dependent random quantity, and the symbol $\corr{\dots}_{d'}$ stands for averaging over the distribution of the diagonal matrix element $d'$ of the Hamiltonian $\widehat H_2$. The expressions provided above refine Eq.\ (\ref{Kom-Lorenz}) at a scale of $\tilde\omega$ of the order of or less than the mean level spacing $\delta$.

Remarkably, the last term of Eq.\ (\ref{gen-R}) is exactly equal to $C(\omega,\lambda)/N$. Thus the global DoS correlation function includes the local one, albeit as a $1/N$-effect.

The formal derivation of the results \eqref{gen-R} and \eqref{gen-C} is given in Appendix \ref{App_sec:NLSM_derivation}.

%%%%%%%%%%%%%%%%%%%%%%%%%%%%%%%%%%%%%%%%%%%%%%%%%%%%%%%%%%%%%%%%%%%%%%%%%%%%%%%%%%%%%%%%%%%%%%%%%%%%%%%%%%%%%%%%%%%
\subsection{DoS correlation function $R(\omega,\lambda)$: smearing of the delta peak}
%%%%%%%%%%%%%%%%%%%%%%%%%%%%%%%%%%%%%%%%%%%%%%%%%%%%%%%%%%%%%%%%%%%%%%%%%%%%%%%%%%%%%%%%%%%%%%%%%%%%%%%%%%%%%%%%%%%

Of special interest is the behavior of $R(\omega,\lambda)$ at small $\omega\ll \Gamma$
since it gives a response of individual levels to perturbation.
In order to obtain an approximate expression in this limit, one can neglect the last term in Eq.\ (\ref{gen-R}) and assume $P_R= 1$. Then expanding $\tilde{M}_{B}^{-1/2}$ in the action and keeping the leading terms in $\omega/\Gamma$ and $\lambda d'$ one obtains after the averaging over $d'$:
\be\label{R-F}
  R(\omega,\lambda)=1+\frac{1}{2} \Re F(x,\alpha),
\ee
where
\be\label{F}
  F(x,\alpha)
  =
  \int_{-1}^{1}d\lambdaF\int_{1}^{\infty}d\lambdaF
  \exp[(ix-\alpha\lambdaB) (\lambdaB-\lambdaF)]
\ee
and we introduced
 two dimensionless parameters:
\begin{gather}
  x = \pi\omega/\delta,
\\
\label{alpha}
\alpha
  =
  \frac{\pi \lambda^2 \corr{d'^2}}{\Gamma\delta}.
\end{gather}
Integration over $\lambdaF$ in Eq.\ (\ref{F}) can be easily performed and we arrive at
\be\label{R-lb}
  R(\omega,\lambda)=1+\Re \! \int_{1}^{\infty}d\lambdaB\,e^{-(\alpha\lambdaB-ix)\lambdaB}
\frac{\sinh(\alpha\lambdaB-ix)}{\alpha\lambdaB-ix}.
\ee
The function $R(\omega,\lambda)$ at various values of the control parameter $\lambda$ is plotted in Fig.\ \ref{fig:CRK}(a).
At $\lambda\rightarrow 0$ this function approaches the Wigner-Dyson limit (shown by the gray dashed line)
\be
  R_\text{WD}(\omega)
  =
  \delta(x)+1-(\sin x/x)^2
\ee
characterized by the $\delta$-function peak of the level self-correlation and quadratic level-repulsion $R_\text{WD}(\omega)\sim x^2$ at $x\ll 1$. At sufficiently small $\lambda$ the self-correlation peak and small-$x$ suppression of $R(\omega,\lambda)$ are still present. However, the peak is progressively broadened and lowered  as $\lambda$ increases and eventually eats up all the correlation hole making the function $R(\omega,\lambda)\approx 1$ essentially flat, see Fig.~\ref{fig:CRK}(a).

%%%%%%%%%%%%%%%%%%%%%%%%%%%%%%%%%%%%%%%%%%%%%%%%%%%%%%%%%%%%%%%%%%%%%%%%%%%%%%%%%%%%%%%%%%%%%%%%%%%%%%%%%%%%%%%%%%%
\subsection{LDoS correlation function $C(\omega,\lambda)$}
\label{SS:C}
%%%%%%%%%%%%%%%%%%%%%%%%%%%%%%%%%%%%%%%%%%%%%%%%%%%%%%%%%%%%%%%%%%%%%%%%%%%%%%%%%%%%%%%%%%%%%%%%%%%%%%%%%%%%%%%%%%%

Comparing Eq.\ \eqref{Kom-Lorenz} with Eqs.\ \eqref{CC} and \eqref{gen-C}, we see that $\tilde K(\omega,\lambda) = C^\text{(L)}(\omega,\lambda)$, where $C^\text{(L)}(\omega,\lambda)$ is the first (Lorentzian) term in the LDoS correlation function.
This term describes the random shift of the \emph{entire mini-band} following the shift of a diagonal term in the Hamiltonian, see Fig.\ \ref{fig:main-result}(a). The coincidence of the two results is an indirect proof of the Monthus surmise, Eq.\ (\ref{Bogomol}).

At an energy scale $\tilde{\omega}\sim \Gamma$ the second term in Eq.\ (\ref{gen-C}) is a correction to the first one at $\tilde{\omega}\sim \Gamma$. However, at $\tilde{\omega}\lesssim \delta\ll \Gamma$ it becomes  significant, as it describes the shift of \emph{a single level} due to perturbation.
In this region Eq.\ (\ref{gen-C}) reduces to
\be\label{C-Phi}
  \bar{C}(\omega,\lambda)=1+\Re\Phi(x,\alpha),
\ee
where, similar to Eq.\ \eqref{F},
\be\label{Phi}
\Phi(x,\alpha)=  \int_{-1}^{1}d\lambdaF\int_{1}^{\infty}
\!
\frac{\lambdaB d\lambdaB}{\lambdaB-\lambdaF} \exp[(ix-\alpha\lambdaB)(\lambdaB-\lambdaF)].
\ee
In order to facilitate the integration in Eq.\ (\ref{Phi}) we first differentiate it with respect to $ix$ and perform integration over $\lambdaF$. Then integration over $ix$ can be done using the formula
\be\label{z-int}
\int e^{-\lambdaB z} \frac{\sinh z}{z}dz
=
\frac{\Ei[z(1-\lambdaF)]-\Ei[-z(1+\lambdaB)]}{2},
\ee
where $z=-ix+\alpha\lambdaB$ and $\Ei(x)$ is the exponential integral function. Finally, one arrives at
\be\label{barC-int}
\bar{C}(\omega,\lambda)
=
1+\Re\int_{1}^{\infty}[I(\lambdaB;x,\alpha)+I(-\lambdaB;-x,\alpha)]\,d\lambdaB,
\ee
where
\be\label{int}
  I(\lambdaB;x,\alpha)=\lambdaB \Ei[ix-\lambdaB(\alpha-ix)-\alpha\lambdaB^{2}].
\ee
The function $\bar{C}(\omega,\lambda)$ is plotted in Fig.\ \ref{fig:CRK}(b) and (c). It also shows a broadened self-correlation peak and the correlation hole, the latter being essentially the same in the LDoS correlation function $\bar{C}(\omega,\lambda)$ and in the global DoS correlation function $R(\omega,\lambda)$.
%%%%%%%%%%%%%%%%%%%%%%%%%%%%%%%%%%%%%%%%%%%%%%%%%%%%%%%%%%%%%%%%%%%%%%%%%%%%%%%%%%%%%%%%%%%%%%%%%%%%%%%%%%%%%%%%%%%
 \begin{figure}[t!]
\center{
\includegraphics[width=0.8 \linewidth,angle=0]{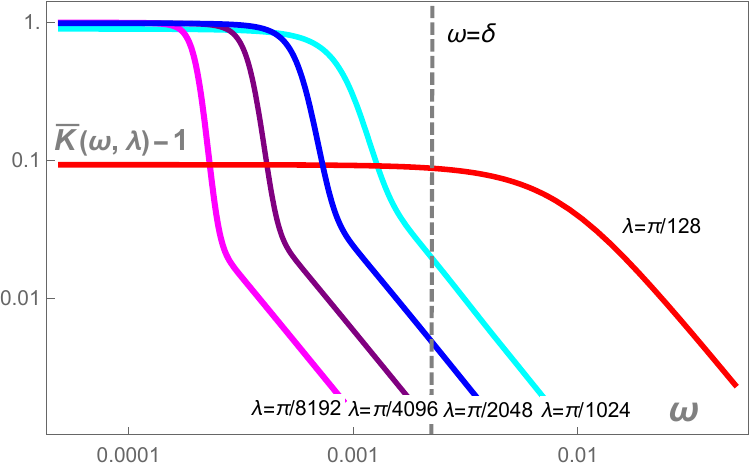}
}
\caption{\textbf{Broadened self-correlation peak}.
The function $ \bar{K}(\omega,\lambda)-1\equiv \bar{C}(\omega,\lambda)/R(\omega,\lambda)-1$ vs.\ $\omega$ for various control parameters $\lambda=\pi/128$, $\pi/1024$, $\pi/2048$, $\pi/4096$, $\pi/8192$ obtained using Eqs.\ (\ref{R-lb}) and (\ref{barC-int}).
The parameters of the GRP model are $N=1024$ and $\gamma=3/2$.
At a small perturbation $\alpha\ll1$ it has almost a box-shaped form with a width
$\sim \alpha^{1/2}\delta$ and a tail $\propto \omega^{-2}$ (the cyan through magenta curves). At large perturbation $\alpha\gg 1$ the peak approaches a Lorentzian shape (the red curve). }
\label{fig:peak}
\end{figure}
%%%%%%%%%%%%%%%%%%%%%%%%%%%%%%%%%%%%%%%%%%%%%%%%%%%%%%%%%%%%%%%%%%%%%%%%%%%%%%%%%%%%%%%%%%%%%%%%%%%%%%%%%%%%%%%%%%%

%%%%%%%%%%%%%%%%%%%%%%%%%%%%%%%%%%%%%%%%%%%%%%%%%%%%%%%%%%%%%%%%%%%%%%%%%%%%%%%%%%%%%%%%%%%%%%%%%%%%%%%%%%%%%%%%%%
\subsection{EAO correlation function $K(\omega,\lambda)$ at small $\lambda$}
%%%%%%%%%%%%%%%%%%%%%%%%%%%%%%%%%%%%%%%%%%%%%%%%%%%%%%%%%%%%%%%%%%%%%%%%%%%%%%%%%%%%%%%%%%%%%%%%%%%%%%%%%%%%%%%%%%

It is remarkable that in the function $K(\omega,\lambda)$ the level repulsion features cancel out [see the inset in Fig.\ \ref{fig:CRK}(b)]. That supports the assumption of the statistical independence of the eigenfunction and spectral fluctuations in the GRP model. The function $\bar{K}(\omega,\lambda)-1\equiv \bar{C}(\omega,\lambda)/R(\omega,\lambda)-1$
is strictly positive, albeit vanishing in the limit $\omega\gg \delta$. Thus the effect of {\it spectral} correlations (the correlation hole due to level repulsion) is absent in $K(\omega,\lambda)$, so that it is essentially  dominated by the {\it eigenfunction amplitude correlations} and describes the broadened self-correlation peak. This peak represents a (not normalized) distribution of the shift of a level at a small perturbation. Figure~\ref{fig:peak} demonstrates that the width of the peak, which gives a typical shift $\Delta E_{n}=E_{n}^{(\lambda)}-E_{n}^{(0)}$ of an individual level, is proportional to $\alpha^{1/2}\delta$
at $\alpha\ll 1$, i.e.\ it is linear in the perturbation strength $\lambda$:
\be\label{width}
\frac{\Delta E_{n}}{\delta}\sim \alpha^{1/2} \sim  \lambda \langle d'^2\rangle^{1/2}\, N^{\gamma/2} \qquad(1<\gamma<2).
\ee

%%%%%%%%%%%%%%%%%%%%%%%%%%%%%%%%%%%%%%%%%%%%%%%%%%%%%%%%%%%%%%%%%%%%%%%%%%%%%%%%%%%%%%%%%%%%%%%%%%%%%%%%%%%%%
\section{Numerical results}
\label{sec: Dos_LDoS}
%%%%%%%%%%%%%%%%%%%%%%%%%%%%%%%%%%%%%%%%%%%%%%%%%%%%%%%%%%%%%%%%%%%%%%%%%%%%%%%%%%%%%%%%%%%%%%%%%%%%%%%%%%%%%%

In this section we present the results of numerical computation of the parametric correlation function $\tilde K(\omega,\lambda)$ of the density operators [Eq.\ (\ref{tilde-K})]. In practice, two narrow bins were located at energies $E=\varepsilon$ and $E=\varepsilon+\omega$ with $\varepsilon$ being close to the band center. In each disorder realization the sum of the products
$\rho_{n}^{(0)}(r)\rho_{m}^{(\lambda)}(r)\equiv \rho_{1}\rho_{2}$ of those states whose energies $E_{n}^{(0)}$ and $E_{m}^{(\lambda)}$ fall in the first and second bins, respectively, was divided by the product of the number of states $n_{1}$ and $n_{2}$ in each bin and {\it after that} averaged over GRP ensemble realizations. The scan over $\varepsilon$ in an energy strip centered at $E=0$ and containing $1/32$ of all levels was also taken to improve statistics.

It is important that all realizations with $n_{1}=0$ and/or $n_{2}=0$ were discarded from the average. For small enough bin widths, $n_{1,2}$ took predominantly only two values (0 or 1), so all realizations with $\{n_{1},{n_{2}}\}=\{1,0\}$, $\{0,1\}$, $\{0,0\}$ were discarded. This procedure eliminated level correlations which would exhibit themselves in the enhanced probability, due to level repulsion, to detect configurations $\{1,0\}$ and $\{0,1\}$ compared to $\{1,1\}$, if all configurations were counted. If only the realizations with $\{1,1\}$ are counted in the ensemble averaging, the level correlation are eliminated by definition.
Note that the function $K(\omega,\lambda)$ defined in Eq.\ (\ref{K_C_R}) would correspond to averaging over all realizations of the sum of the products $\rho_{1}\rho_{2}$ divided by the average of $n_{1}n_{2}$ over all realizations.

The calculations of this section involve the numerical diagonalization of random matrices from the GRP ensemble of the \emph{orthogonal} symmetry, in contrast to the analytical results of Sec.\ \ref{sec: Dos_LDoS-analyt} derived for the \emph{unitary} GRP. The similarity of the numerical results for $\tilde{K}(\omega,\lambda)$ to those derived analytically for $K(\omega,\lambda)$ corroborates the assumption of statistical independence of eigenfunctions and eigenvalues in GRP ensemble and demonstrates that there is no qualitative difference in the eigenfunction overlap statistics caused by the difference in the
 symmetry class.

%%%%%%%%%%%%%%%%%%%%%%%%%%%%%%%%%%%%%%%%%%%%%%%%%%%%%%%%%%%%%%%%%%%%%%%%%%%%%%%%%%%%%%%%%%%%%%%%%%%%%%%%%%%%%%%%
\subsection{Diagonal drive}
%%%%%%%%%%%%%%%%%%%%%%%%%%%%%%%%%%%%%%%%%%%%%%%%%%%%%%%%%%%%%%%%%%%%%%%%%%%%%%%%%%%%%%%%%%%%%%%%%%%%%%%%%%%%%%%%%

Here we present the results of numerical diagonalization of matrices described by Eq.\ (\ref{H-1-2}), where $\hat{H}_{1}$ is drawn from the GRP ensemble governed by Eq.\ (\ref{variances}) and $\hat{H_{2}}$ is an independently of ${\hat{H}_{1}}$ fluctuating diagonal random matrix 
$\hat{H_{2}}=\diag\{d'_{n}\}$ with the same variance of diagonal elements: $\corr{d_n'^2}=W^2$.

%%%%%%%%%%%%%%%%%%%%%%%%%%%%%%%%%%%%%%%%%%%%%%%%%%%%%%%%%%%%%%%%%%%%%%%%%%%%%%%%%%%%%%%%%%%%%%%%%%%%%%%%%%%%%%%%%%%
\begin{figure}[tb]
\center{
\includegraphics[width=0.8 \linewidth,angle=0]{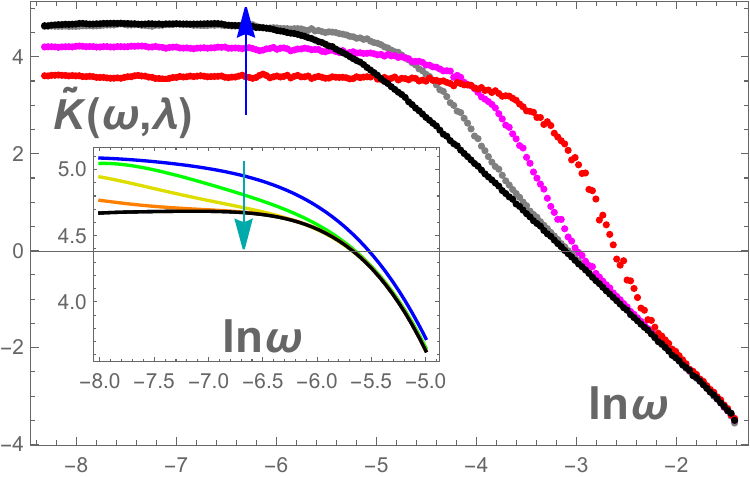}
}
\caption{\textbf{EAO for the diagonal drive}. The function $\tilde{K}(\omega,\lambda)$ from the numerical diagonalization of the Hamiltonian (\ref{H-1-2}) with a diagonal $\hat{H}_{2}$ at
$N=1024$, $W=1$, $\sigma^{2}=0.1\,N^{-\gamma}$, $\gamma=3/2$ and $\lambda=\pi/512$ (gray), $\pi/256$ (magenta), $\pi/128$ (red) and $\lambda=0$ (black). The arrows show the direction of evolution as $\lambda$ is decreasing. \textbf{Inset}: the broadened self-correlation bump in $\tilde{K}(\omega,\lambda)$ at small $\omega$ for a weak drive $\lambda=\pi/2048$, (blue) $\lambda=\pi/4096$ (green), $\lambda=\pi/8192$ (yellow), $\lambda=\pi/16384$ (orange).   The self-correlation peak makes the direction of evolution opposite to that at a strong drive on the main plot.  }
\label{fig:Kom-diag}
\end{figure}
%%%%%%%%%%%%%%%%%%%%%%%%%%%%%%%%%%%%%%%%%%%%%%%%%%%%%%%%%%%%%%%%%%%%%%%%%%%%%%%%%%%%%%%%%%%%%%%%%%%%%%%%%%%%%%%%%%%
%%%%%%%%%%%%%%%%%%%%%%%%%%%%%%%%%%%%%%%%%%%%%%%%%%%%%%%%%%%%%%%%%%%%%%%%%%%%%%%%%%%%%%%%%%%%%%%%%%%%%%%%%%%%%%%%%%%
\begin{figure}[t]
\center{
\includegraphics[width=0.8 \linewidth,angle=0]{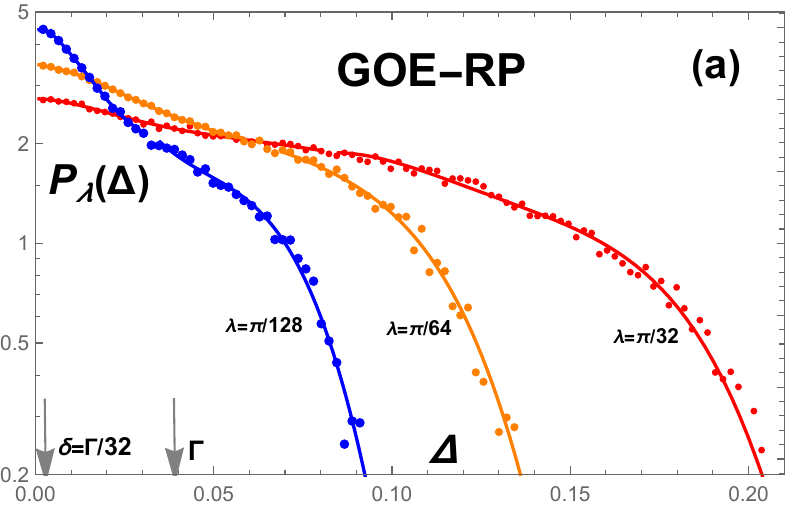}
\includegraphics[width=0.8 \linewidth,angle=0]{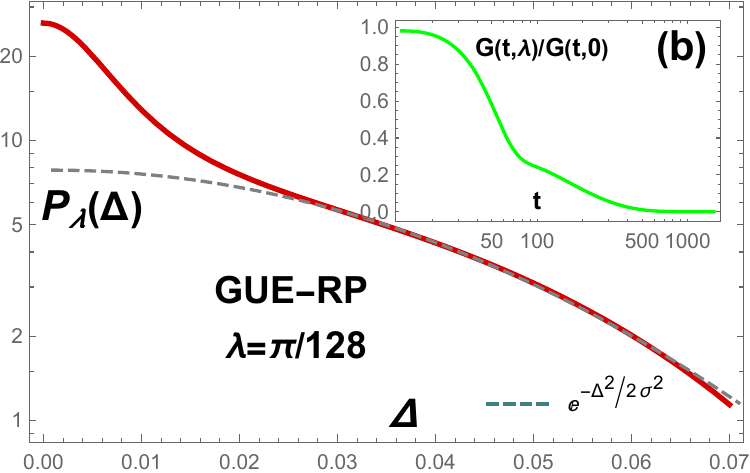}
}
\caption{\textbf{Distribution of shifts in the local spectrum, ${\cal P}_{\lambda}(\Delta)$}.
\textbf{(a)} The result of numerical diagonalization of matrices
from the orthogonal RP ensemble with
$N=1024$, $W=1$, $\sigma^{2}=N^{-\gamma}$, $\gamma=3/2$ and $\lambda=\pi/128$, $\pi/64$, $\pi/32$. The Fourier transform $G(t,\lambda)$ of the EAO correlation function $\tilde{K}(\omega,\lambda)$ expressed in terms of eigenvectors and eigenvalues was computed directly and  Fourier-transformed according to Eq.\ (\ref{P-Delta}).
\textbf{(b)} The analytical result obtained by Fourier transforming $K(\omega,\lambda)$ given by Eqs.\ (\ref{K_C_R}), (\ref{gen-R}), (\ref{gen-C}) and inverse Fourier transforming according to Eq.\ (\ref{P-Delta}). The function ${\cal P}_{\lambda}(\Delta)$ consists of a  narrow peak at small $\Delta$ [which reflects the broadened self-correlation peak in $K(\omega,\lambda)$, see the red curve in Fig.\ \ref{fig:peak}] on the top of a much broader distribution of the shifts of a mini-band as a whole, see Fig.\ \ref{fig:main-result}(a). The dashed line shows the Gaussian fit to this distribution. The similar features are seen in the numerical result (blue curve on the upper panel).
\textbf{Inset}: The ratio $G(t,\lambda)/G(t,0)$ whose inverse Fourier transform gives ${\cal P}_{\lambda}(\Delta)$.  }
\label{fig:P}
\end{figure}
%%%%%%%%%%%%%%%%%%%%%%%%%%%%%%%%%%%%%%%%%%%%%%%%%%%%%%%%%%%%%%%%%%%%%%%%%%%%%%%%%%%%%%%%%%%%%%%%%%%%%%%%%%%%%%%%%%

In Fig.\ \ref{fig:Kom-diag} we plot the function $\tilde{K}(\omega,\lambda)$ obtained for $N=1024$, $W=1$, $\sigma^2=0.1 N^{-\gamma}$, $\gamma=3/2$ and different values of $\lambda$.
At a strong drive (the main plot) the results of numerical diagonalization are very similar to the prediction of the Monthus surmise presented in Fig.~\ref{fig:K-Gamma}.
At a weak drive the evolution of $\tilde{K}(\omega,\lambda)$ at small (fixed) $\omega$ is opposite to that of the strong drive due to the broadened self-correlation peak
 whose
 front is shifted to smaller values of $\omega$ as $\lambda$ decreases (see Fig.\ \ref{fig:peak}).

It is instructive to extract the effective distribution ${\cal P}_{\lambda}(\Delta)$ of level shifts $\Delta$ in the LDoS defined by Eq.\ (\ref{P-Delta}). At the scale $\Delta\gtrsim \Gamma$ this function gives the distribution of random shifts of the entire mini-band in the LDoS, while at $\Delta\lesssim \delta$ it gives the distribution of random shifts of an individual level. To find ${\cal P}_{\lambda}(\Delta)$ we compute numerically the Fourier transform $G(t,\lambda)$ of $\tilde{K}(\omega,\lambda)$ and apply Eq.\ (\ref{P-Delta}) performing the discrete (inverse) Fourier transform. The result is presented in Fig.\ \ref{fig:P} together with the corresponding analytical result obtained from Eqs.\ (\ref{K_C_R}), (\ref{gen-R}) and (\ref{gen-C}).
Note however that the analytical result is obtained for the unitary symmetry, while the numerics is done for the orthogonal symmetry, so only a qualitative agreement between Fig.\ \ref{fig:P}(a) and Fig.\ \ref{fig:P}(b) is expected.

%%%%%%%%%%%%%%%%%%%%%%%%%%%%%%%%%%%%%%%%%%%%%%%%%%%%%%%%%%%%%%%%%%%%%%%%%%%%%%%%%%%%%%%%%%%%%%%%%%%%%%%%%%%%%%%%%%%
\begin{figure}[t]
\center{
\includegraphics[width=0.8 \linewidth,angle=0]{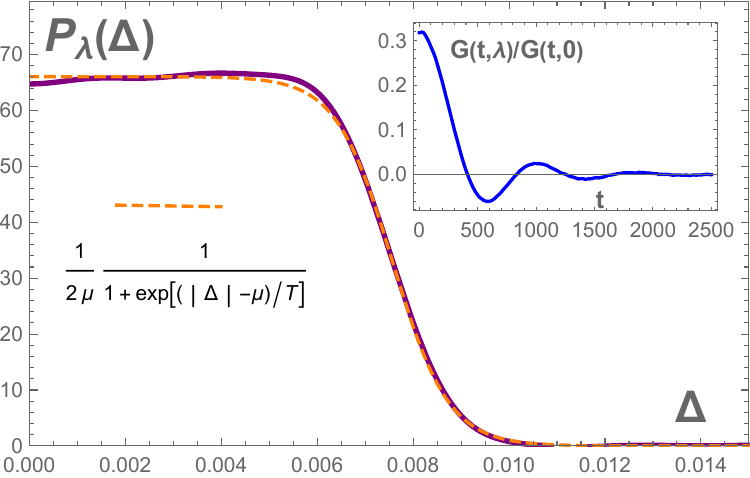}
}
\caption{\textbf{${\cal P}_{\lambda}(\Delta)$ for the off-diagonal drive}.   The result of numerical diagonalization
for the orthogonal RP ensemble with
$N=1024$, $W=1$, $\sigma^{2}=N^{-\gamma}$, $\gamma=3/2$ and $\lambda=\pi/128$.
The solid purple curve is the Fourier transform according to Eq.\ (\ref{P-Delta}) of $G(t,\lambda)/G(t,0)$ found from exact diagonalization and shown in the inset. The dashed orange line is the fit to the Fermi-like function \eqref{Fermi} with $\mu=0.00757$ and $T=0.000585$.}
\label{fig:P_off}
\end{figure}
%%%%%%%%%%%%%%%%%%%%%%%%%%%%%%%%%%%%%%%%%%%%%%%%%%%%%%%%%%%%%%%%%%%%%%%%%%%%%%%%%%%%%%%%%%%%%%%%%%%%%%%%%%%%%%%%%%

\subsection{Off-diagonal drive}

%%%%%%%%%%%%%%%%%%%%%%%%%%%%%%%%%%%%%%%%%%%%%%%%%%%%%%%%%%%%%%%%%%%%%%%%%%%%%%%%%%%%%%
When the diagonal matrix elements in Eq.\ (\ref{H-1-2}) do not move, i.e.\ $\hat{H}_{2}$ is an off-diagonal part of a matrix drawn from the GRP ensemble and $\hat{H}_{1}$ is an independently fluctuating full matrix from the same ensemble, the distribution function ${\cal P}_{\lambda}(\Delta)$ looks differently from the
one at the diagonal drive. First of all, there is no shift of the mini-band as a whole that manifests itself by the absence of the broad part of ${\cal P}_{\lambda}(\Delta)$ in Fig.\ \ref{fig:P_off}.
Remarkably, the shape of ${\cal P}_{\lambda}(\Delta)$ [obtained by Fourier transforming $G(t,\lambda)/G(t,0)$ found by direct numerical calculations, see the inset to Fig.\ \ref{fig:P_off})] has the form of the Fermi distribution:
\be\label{Fermi}
{\cal P}_{\lambda}(\Delta)= \frac{C}{1+\exp[(|\Delta|-\mu)/T]},
\ee
where $C^{-1}=2T\ln(1+e^{\mu/T})\approx 2\mu$.

On the main plot of Fig.\ \ref{fig:P_off} the numerical result for ${\cal P}_{\lambda}(\Delta)$ is shown by a purple curve and the orange dashed line is the fit by Eq.\ (\ref{Fermi}).
As one can see, the fit is almost perfect. As $\lambda$ decreases, the width $\mu$ of the distribution and the ``effective temperature'' $T$ both decrease but the ratio $T/\mu$ remains small and is likely to vanish in the limit $\lambda\rightarrow 0$. This behavior is similar to the one obtained analytically for the diagonal drive and shown in Fig.\ \ref{fig:peak}, but in contrast to the diagonal drive, the Lorentzian tail is absent at the off-diagonal drive.

%%%%%%%%%%%%%%%%%%%%%%%%%%%%%%%%%%%%%%%%%%%%%%%%%%%%%%%%%%%%%%%%%%%%%%%%%%%%%%%%%%%%%%%%%%%%%%%%%%%%%%%%%%%%%%%%%%%
\begin{figure}[t!]
\center{
\includegraphics[width=0.8\linewidth]{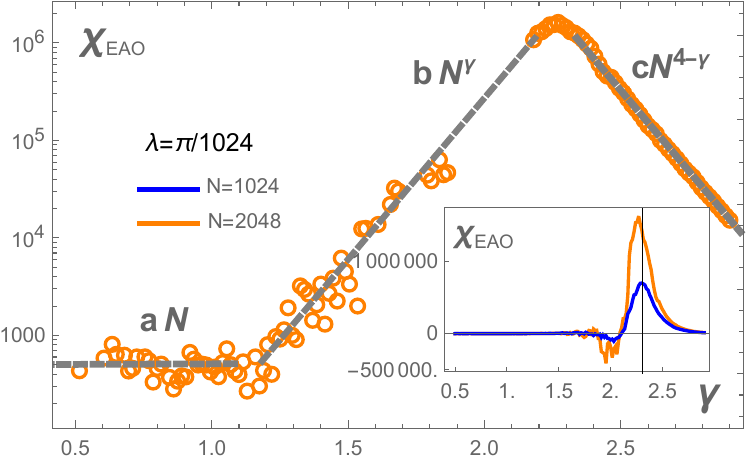}
}
\caption{\textbf{Eigenfunction overlap susceptibility} defined by Eq.\ (\ref{EAOS}) and obtained from the numerical diagonalization as a function of $\gamma$. Its scaling with $N$ is determined by Eqs.\ (\ref{chi-chi}) and (\ref{scaling-K-delta}), as indicated on the plot.
 The shift of the maximum away from $\gamma_\text{AT}=2$ is a finite-size effect (see text).
\textbf{Inset}: $\chi_\text{EAO}$ in a linear scale for $N=1024$ and $N=2048$, showing a shift of the maximum closer to $\gamma_\text{AT}$ with increasing $N$.}
\label{fig:chi_EAOS}
\end{figure}
%%%%%%%%%%%%%%%%%%%%%%%%%%%%%%%%%%%%%%%%%%%%%%%%%%%%%%%%%%%%%%%%%%%%%%%%%%%%%%%%%%%%%%%%%%%%%%%%%%%%%%%%%%%%%%%%%%

In order to estimate the expected width of the distribution ${\cal P}_{\lambda}(\Delta)$ one can use
Eq.\ (\ref{alpha}) derived for the diagonal drive in which the ``driving force'', the variance $\langle d'^{2}\rangle$, should be replaced by
$\sum_{m}\langle (H_{2,nm})^{2}\rangle
= N \sigma^2 \propto N^{1-\gamma}$.
Then the parameter $\alpha$ transforms to
\be
\label{alpha_off}
  \alpha_\text{off}=\lambda^2N/2,
\ee
and using Eq.\ (\ref{width}) we obtain
\be
\Delta E_{n}\equiv\mu \sim \delta\, (\lambda \sqrt{N})\sim \lambda W N^{-1/2}.
\ee
Hence the typical shift of the level $\mu$ for the off-diagonal drive in the fractal phase ($1<\gamma<2$) is independent of~$\gamma$.

Finally, we present the numerical results for the sensitivity of the EAO correlation function $\delta \tilde{K}(\delta,\lambda)$ at a scale $\omega=\delta$ to the change of the control parameter $\lambda$ for the off-diagonal drive. To this end we define the EAO susceptibility according to
\be\label{EAOS}
\chi_\text{EAO} = \frac{\tilde{K}(\delta,\lambda)-\tilde{K}(\delta,0)}{\lambda^2}.
\ee
Figure \ref{fig:chi_EAOS} demonstrates that it has the same shape as the fidelity susceptibility shown in Fig.\ \ref{fig:main-result}(b) and its scaling with the matrix size $N$ differs only by a $\gamma$-independent extra factor of $N$.
This extra factor can be easily understood if we evaluate the numerator in Eq.\ \eqref{EAOS}
expanding the functions $F(\pi,\alpha)$ [Eq.\ \eqref{F}] and $\Phi(\pi,\alpha)$ [Eq.\ \eqref{Phi}] to the first order in $\alpha$ and replacing the latter by $\alpha_\text{off}$:
\be
K(\delta,\lambda)-K(\delta,0) \sim K(\delta,0)\,\alpha_\text{off}
\sim \chi_\text{F}^\text{(typ)}\, \alpha_\text{off} .
\ee
Now taking $\alpha_\text{off}$ from Eq.\ \eqref{alpha_off}, we obtain
\be
\label{chi-chi}
\chi_\text{EAO} \sim N \chi_\text{F}^\text{(typ)} .
\ee
This similarity (confirmed numerically in Fig.\ \ref{fig:chi_EAOS})
implies that not only the leading in $\lambda^2$ correction to the fidelity, $1-F^2\approx\lambda^2\chi_\text{F}^\text{(typ)}$, but also the next in $\lambda^2$ correction,
$\lambda^2[K(\delta,\lambda)-K(\delta,0)]=\lambda^4\chi_\text{EAO}$ as functions of $\gamma$ are peaked at the localization transition.
The apparent shift of the maximum $\gamma_\text{max}$ away from the transition point $\gamma_\text{AT}=2$ is a finite-size effect. Indeed, with the constants $b,c\sim 1$ defined in Fig.\ \ref{fig:chi_EAOS}, one finds a correction $(1/2)\ln(c/b)/\ln N\sim 15\%$ to $\gamma_\text{max}$.
A similar correction would appear for $\gamma_{\text{max}}$ in $\chi_\text{F}^\text{(typ)}$ if the corresponding prefactors $\tilde{b}$ and $\tilde{c}$ are taken into account in Eq.\ (\ref{scaling-K-delta}). Taking the log-derivative as in Eq.\ (\ref{log-der-chi}) eliminates this correction and brings $\gamma_{\text{max}}$ almost exactly to $\gamma_{\text{AT}}$, see Fig.\ \ref{fig:main-result}(b). Unfortunately, strong data scattering makes this operation impossible for $\chi_\text{EAO}$.

Equation (\ref{chi-chi}) may be considered as a numerical justification of the replacement $\alpha\to\alpha_\text{off}$, with $\alpha_\text{off}$ given by Eq.\ (\ref{alpha_off}), for the small-$\lambda$ expansion of the fidelity $F$ in the case of the off-diagonal drive.

\begin{figure}[t]
\center{
\includegraphics[width=0.8 \linewidth,angle=0]{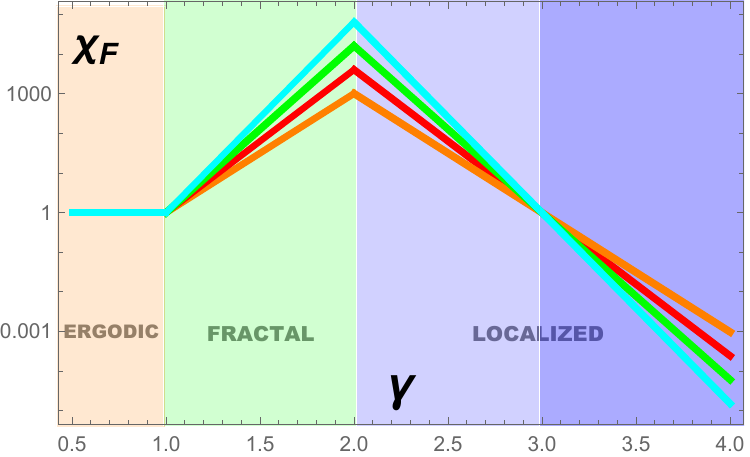}
}
\caption{\textbf{Sketch of the fidelity susceptibility} as a function of the effective disorder parameter $\gamma$ at different system sizes (from orange to cyan in increasing $N$ order) in the thermodynamic limit, $N\to\infty$. For finite $N$, $\chi_{\text{F}}^{\text{typ}}$ acquires $\gamma$-dependent corrections rounding the curves and shifting the maximum, as in Fig.\ \ref{fig:chi_EAOS}.
The plot corresponds to Fig.\ \ref{fig:main-result}(b) and should be compared with Fig.\ 1 in Ref.\ \cite{Sels21}.}
\label{fig:chi-sketch}
\end{figure}

%%%%%%%%%%%%%%%%%%%%%%%%%%%%%%%%%%%%%%%%%%%%%%%%%%%%%%%%%%%%%%%%%%%%%%%%%%%%%%%%%%%%%%%
\section{Conclusion }\label{sec:Conclusion}
%%%%%%%%%%%%%%%%%%%%%%%%%%%%%%%%%%%%%%%%%%%%%%%%%%%%%%%%%%%%%%%%%%%%%%%%%%%%%%%%%%%%%%%

In this paper we studied the response of an isolated quantum system governed by the Gaussian Rosenzweig-Porter random matrix Hamiltonian $\hat{H}_1$ to a perturbation $\lambda\hat{H}_2$.
In contrast to the disordered spin chains, where a complete analytical treatment was not possible so far, our random matrix model can be exactly solved in the region of ergodic and fractal states  by the formalism of the Efetov's supersymmetric nonlinear sigma model in the thermodynamic limit $N\rightarrow\infty$. This solution has a natural extension  to the region of localized states which is successfully tested numerically. 

In this work we obtained asymptotically exact expressions, Eqs.\ (\ref{gen-R}) and (\ref{gen-C}), for the DoS and LDoS correlation functions in the GRP ensemble of unitary symmetry in its fractal phase. Having calculated the eigenfunction amplitude overlap correlation function given by Eq.\ \eqref{K_C_R}, we compared it with simple analytical treatment based on the Monthus surmise [Eq.\ (\ref{Bogomol})], as well as with the numerical simulations on the orthogonal-symmetry version of the same ensemble. The studied correlation functions behave qualitatively in the same way for both symmetries. We identified analytically the parameter $\alpha$ [Eqs.\ \eqref{alpha} and \eqref{alpha_off} for different types of drive], which controls a typical shift of individual levels by a small perturbation of the Hamiltonian.

We focus on the typical fidelity susceptibility $\chi_\text{F}^\text{(typ)}$ [Eq.\ (\ref{chiF})] and the susceptibility of the eigenfunction amplitude overlap $\chi_\text{EAO}$ [Eqs.\ (\ref{EAOS}) and (\ref{tilde-K})] to such a perturbation. We show that $\chi_\text{F}^\text{(typ)}$ and $\chi_\text{EAO}$ contain essentially the same information on the effect of the perturbation on the eigenfunction overlap.

It is demonstrated both analytically and numerically that these susceptibilities as functions of the effective disorder parameter $\gamma$ are strongly peaked near the localization transition: they are nearly constant in the ergodic phase, increase exponentially in the fractal phase and decrease exponentially in the localized phase. Furthermore, in the localized phase we identified two drastically different regimes of the  behavior of $\chi_\text{F}^\text{(typ)}\sim \chi_\text{EAO}/N$ with increasing the matrix size $N$ (which is the dimension of the Hilbert space in this problem). Both susceptibilities grow with $N$ for $2<\gamma<3$ and decrease with $N$ for $\gamma>3$, with a fixed point at $\gamma^{*}=3$.

The overall behavior of $\chi_\text{F}^\text{(typ)}$ sketched in Fig.\ \ref{fig:chi-sketch} is very similar to the one for the scaled typical normalized fidelity susceptibility of a true many-body system, the disordered XXZ spin chain, studied recently by Sels and Polkovnikov \cite{Sels21}.
This similarity includes also the existence of an apparent fixed point that separates the regimes of increasing and decreasing $\chi_\text{F}^\text{(typ)}$ with the Hilbert space dimension $N$.
We argue that the first regime is realized when two resonant localized states may hybridize and form the Mott's pair of bonding and anti-bonding states with the level splitting $\Delta E\sim\delta$. The second regime implies that such ``double-headed'' states are rare and make a subleading contribution compared to typical ``single-headed'' ones.

We believe that the similarity of our Fig.\ \ref{fig:main-result}(b) and Fig.\ 1 of Ref.\ \cite{Sels21} is not accidental, reflecting the common physics both in the many-body problem of disordered spin chains and in the RP random matrix problem.

\begin{acknowledgments}

We are grateful to G. De Tomasi and  I. M. Khaymovich for multiple illuminated discussions, and specifically to I. M. Khaymovich for the help in numerical calculations and  for allowing us to use the data obtained in the course of work on Ref.\ \cite{LN-RP-RRG21} that led to the plot in Fig.~\ref{fig:main-result}(c). We appreciate the contribution of M. V. Feigel'man in formulating the problem and interest to the work.

We acknowledge the CINECA award under the ISCRA initiative, for the availability of high performance computing resources and support.
The support from ICTP Associate Program (M.A.S.) and from Google Quantum Research Award ``Ergodicity breaking in Quantum Many-Body Systems'' (V.E.K.) is gratefully acknowledged.

\end{acknowledgments}

\appendix

%%%%%%%%%%%%%%%%%%%%%%%%%%%%%%%%%%%%%%%%%%%%%%%%%%%%%%%%%%%%%%%%%%%%%%%%%%%%%%%%%%%%%%

\section{DERIVATION OF EQS.\ (\ref{gen-R}), (\ref{gen-C}) BY~THE EFETOV'S NLSM FORMALISM}
\label{App_sec:NLSM_derivation}

%%%%%%%%%%%%%%%%%%%%%%%%%%%%%%%%%%%%%%%%%%%%%%%%%%%%%%%%%%%%%%%%%%%%%%%%%%%%%%%%%%%%%%

In this Appendix we outline the main steps in the derivation of the analytical results (\ref{gen-R}) and (\ref{gen-C}) for the DoS and LDoS parametric correlation functions in the fractal phase of the unitary GRP model subject to the diagonal drive. The details will be reported elsewhere~\cite{2B-SK}.

We take the variance of off-diagonal elements of the Hamiltonian in Eq.\ \eqref{variances} to be $\sigma^2=1/N$. Then the fractal nonergodic phase is realized for $1\ll W^2\ll N$. In terms of the energy scales, that corresponds to the hierarchy $\delta \ll \Gamma \ll E_\text{BW} \sim W$, see Table \ref{T:GRP}.

To simplify calculations it is convenient to write the GRP Hamiltonian as a sum of a totally basis-invariant RMT part and a diagonal contribution:
\be
\label{H-new}
  H(\lambda)
  =
  H_\text{RMT} + \hat d \cos\lambda + \hat d' \sin\lambda ,
\ee
where $H_\text{RMT}$ is a random GUE matrix with the variance of all (including diagonal) elements $\corr{|H_{nm}|^2}=1/N$, while $\hat d=\diag(d_n)$ and $\hat d'=\diag(d'_n)$ are diagonal matrices independently distributed with zero mean and $\corr{d_n^2}=\corr{d_n^{\prime2}}=W^2-1/N$ (since the variance of a sum of normally distributed random variables is additive). In the fractal phase the last ($1/N$) term is negligible that will be assumed hereafter.

In the self-consistent Born approximation, justified in the fractal phase since $\Gamma\gg\Delta$,
the average Green function is given by
\be
  \corr{G^\text{R,A}_{ij}(E,\lambda)}
  =
  \frac{\delta_{ij}}{E\pm i\Gamma/2} ,
\ee
where the width is obtained by the Fermi golden rule:
$\Gamma = 2 \pi \corr{|H_{ij}|^2}/\delta$, in accordance with Table \ref{T:GRP}.

The DoS and LDoS correlation functions [Eqs.\ \eqref{R} and \eqref{C}] can be expressed in terms of the average product of the retarded and advanced Green's functions calculated at different values of the control parameter $\lambda$:
\be
\label{Y-def}
  Y_{rs}(\omega,\lambda)
  =
  \corr{G^\text{R}_{rr}(E+\omega/2,-\lambda/2) G^\text{A}_{ss}(E-\omega/2,\lambda/2)} .
\ee
In the RP model, the matrix $Y_{rs}$ is characterized by two parameters: its diagonal ($Y_{rr}$) and off-diagonal ($Y_{r\neq s}$) elements. They determine the correlators in question:
\begin{gather}
  R(\omega,\lambda)
  =
  \frac12 + \frac{\delta^2}{2\pi^2}
  \Re
  \bigl[ N(N-1)Y_{r\neq s} + NY_{rr} \bigr]
\\
  C(\omega,\lambda)
  =
  \frac12 + \frac{N^2\delta^2}{2\pi^2}
  \Re Y_{rr}
\end{gather}

To perform the average in Eq.\ \eqref{Y-def}, we follow the standard Efetov's formalism of the supersymmetric NLSM \cite{Efetov_book}, repeating the sequence of routine steps: representation of the Green's function via integrals over supervectors $\psi$, averaging over the GUE matrix $H_\text{RMT}$, decoupling the resulting quartic-in-$\psi$ term by the Hubbard-Stratonovich transformation that introduces a $4\times4$ supermatrix $Q$ acting in the tensor product of the superspace (FB) and retarded-advanced (RA) space, and taking the final Gaussian integral over $\psi$. We also add a symmetry-breaking source term to the action, $J=\diag\{J_i\}$, where
$J_i=\diag(\xi^\text{R}_i, \xi^\text{A}_i) \otimes \hat P_\text{B}$ and $\hat P_\text{B}$ projects onto the BB sector, and define the partition function
\be
  Z[J] = \int e^{S[Q,J]} DQ .
\ee
Then the correlator \eqref{Y-def} is expressed as
\be
\label{Y-via-Z}
  Y_{rs}
  =
  \frac{\partial^2Z[J]}{\partial\xi^\text{R}_r\partial\xi^\text{A}_s}
  \bigg|_{\xi=0}
  .
\ee

A new ingredient in this standard RMT scheme arising for the RP model is the presence of diagonal disorder in Eq.\ \eqref{H-new} that requires additional averaging. However, since after the Hubbard-Stratonovich transformation different components of the supervector $\psi_i$ ($i=1,\dots,N$) are already decoupled and both disorder ($\hat d$ and $\hat d'$) and the source $J$ are also diagonal, each component can still be averaged independently.
The action then is given by%
\be
\label{action-R-Z}
  S[Q,J]
  =
  - \frac{N\Gamma^2}{8} \str Q^2
  + S_\sigma[Q,J]
\ee
where the last term is obtained as a sum of $N$ independent contributions (assuming the band center, $E=0$, and linearizing for small $\lambda$):
\be
\label{action-sigma-Z}
  S_\sigma[Q,J]
  =
  \sum_{i=1}^N \ln \corr{ \sdet(\tilde\omega\Lambda/2-d+i\Gamma Q/2+J_i) }
  _{d,d'} .
\ee
Here averaging is performed over two independent Gaussian variables $d$ and $d'$ (initially, $i$'th components of $\hat d$ and $\hat d'$), both with the variance $W^2$. Note that $d'$ responsible for the drive enters the action \eqref{action-sigma-Z} via a frequency shift%
\be
  \tilde\omega = \omega-\lambda d' ,
\ee
as follows from the structure of the arguments of the Green's functions in Eq.\ \eqref{Y-def}.

At $\omega=0$, $\lambda=0$ and in the absence of sources ($J=0$), the action \eqref{action-R-Z} does not involve any nontrivial matrices rather than $Q$ ($d$ enters with the unit matrix in the $\text{FB}\otimes\text{RA}$ space) and therefore possesses a degenerate manifold of saddle points $Q=U^{-1}\Lambda U$ spanning the coset $U(2,2)/U(1|1)\times U(1|1)$. The action of the resulting NLSM is given by Eq.\ \eqref{action-sigma-Z}.

Since we are interested in the parametric statistics at small values of the control parameter $\lambda\ll1$, the roles of $d$- and $d'$-averaging are essentially different. The former provides the leading contribution responsible for the formation of the mini-band, whereas the latter describes parametric correlations in the spectrum.

In the absence of sources, the NLSM action \eqref{action-sigma-Z} depends only on Cartan variables, coinciding with $\lambdaF$ and $\lambda_B$ in the Efetov's parametrization \cite{Efetov_book}.
Then
\be
\label{sdet-via-M}
  \sdet(\omega\Lambda/2-d+i\Gamma Q/2)
  =
  \frac
  {4d^2+\tilde M_\text{F}}
  {4d^2+\tilde M_\text{B}} ,
\ee
where we introduced a short-hand notation
\be
\label{MM}
  \tilde M_\text{F,B}
  =
  \Gamma^2 - \tilde \omega^2 - 2 i \Gamma \tilde \omega \lambda_\text{F,B}
  .
\ee
Averaging the superdeterminant over $d$ we obtain the source-free action $S \equiv S[Q,0]$ given by Eq.\ \eqref{action-final}.

Calculation of the derivatives in Eq.\ \eqref{Y-via-Z} is performed in the Efetov's parametrization, which contains four real variables (two Cartan coordinates, $\lambda_\text{F,B}$, and two angular variables) and four Grassmann numbers. After some algebra \cite{2B-SK} we obtain in the leading order in $N\to\infty$:
\begin{gather}
\label{Yrs}
  Y_{r\neq s}
  =
  \frac{\pi^2}{N^2\delta^2}
  \int DQ
  \left[
  1
  +
  (\lambdaB-\lambdaF)^2 P_R G_4
  \right]
  e^{S} ,
\\
\label{Yrr}
  Y_{rr}
  =
  \frac{2\pi}{N\delta\Gamma}
  \int DQ
  \left[
  \Corr{\frac{\Gamma}{\Gamma-i\tilde\omega}}_{\!\!d'}
  +
  (\lambdaB-\lambdaF) P_C G_4
  \right]
  e^{S} ,
\end{gather}
where angular brackets stand for averaging over $d'$,
the functions $P_R(\lambdaB)$ and $P_C(\lambdaB)$ defined in Eqs.\ \eqref{PR} and \eqref{PC} also involve averaging over $d'$,
and $G_4$ is the product of all four Grassmann variables.

The invariant measure for integration over the manifold of $Q$ matrices is given by \cite{Efetov_book}
\be
\label{DQ}
  \int DQ \dots
  =
  \int_{-1}^{1} d\lambdaF
  \int_{1}^{\infty} d\lambdaB
  \frac{dG}{(\lambdaB-\lambdaF)^2}
  \dots ,
\ee
where $dG$ stands for the measure over four Grassmann variables (we omit integration over two angular coordinates since the integrand does not depend on them).
After integration over $dG$ in Eqs.\ \eqref{Yrs} and \eqref{Yrr} the term $G_4$ gives unity, various combinations of two Grassmann variables vanish (therefore they are omitted in those equations), while the Grassmann-free terms do contribute due to the singularity of the measure at $\lambdaF=\lambdaB=1$
(Efetov-Wegner boundary term \cite{Efetov_book}).
Collecting altogether we arrive at Eqs.\ (\ref{gen-R}) and (\ref{gen-C}).

%%%%%%%%%%%%%%%%%%%%%%%%%%%%%%%%%%%%%%%%%%%%%%%%%%%%%%%%%5
\bibliography{parametric-RP-cm-2}
%%%%%%%%%%%%%%%%%%%%%%%%%%%%%%%%%%%%%%%%%%%%%%%%%%%%%%%%%5

\end{document}